\documentclass
[prc,a4paper,twocolumn,showpacs,preprintnumbers,superscriptaddress,floatfix]{revtex4}%
\usepackage{amssymb}
\usepackage{graphicx}
\usepackage{dcolumn}
\usepackage{bm}
\usepackage{longtable}
\usepackage{amsmath}
\usepackage{amsfonts}
\usepackage{booktabs}
\usepackage{dsfont}%
\providecommand{\U}[1]{\protect\rule{.1in}{.1in}}

\begin{document}
\title{Relativistic RPA in axial symmetry}
\author{D. Pena Arteaga$^{1}$ and P. Ring}
\affiliation{Physikdepartment, Technische Universit\"{a}t M\"{u}nchen, D-85748, Garching, Germany}
\affiliation{Departamento de F\'{\i}sica Te\'{o}rica, Universidad Aut\'{o}noma de Madrid,
E-28049 Madrid, Spain}
\date{\today}

\begin{abstract}
Covariant density functional theory, in the framework of self-consistent
Relativistic Mean Field (RMF) and Relativistic Random Phase approximation
(RPA), is for the first time applied to axially deformed nuclei. The fully
self-consistent RMF+RRPA equations are posed for the case of axial symmetry
and non-linear energy functionals, and solved with the help of a new parallel
code. Formal properties of RPA theory are studied and special care is taken in
order to validate the proper decoupling of spurious modes and their influence
on the physical response. Sample applications to the magnetic and electric
dipole transitions in $^{20}$Ne are presented and analyzed.

\end{abstract}

\maketitle

\section{Introduction}

New experimental facilities with radioactive nuclear beams have stimulated
enhanced experimental and theoretical efforts to understand the structure of
nuclei, not only along the narrow line of stable isotopes, but also in areas
of large neutron- and proton excess far from the valley of $\beta$-stability.
Beside the investigation of the ground state properties of these nuclei, more
and more experimental studies are being devoted to the understanding of the
properties of excited states in this region.

On the theoretical side, only very light nuclei can be studied in the
framework of modern \textit{ab initio} me\-thods. Shell model calculations in
restricted configuration spaces provide an accurate description of light and
medium-heavy nuclei. For the large majority of nuclei, however, a quantitative
microscopic description is only possible using density functional theory
(DFT). Although DFT can, in principle, provide an exact description of the
many-body dynamics if the exact density functional is known~\cite{HK.64,KS.65}%
, in nuclear physics one is far from a microscopic derivation of this
functional and in addition there is the problem, that in self-bound systems
density functional theory can only be applied to intrinsic
densites~\cite{Eng.07,Gir.08}. The most successful schemes use a
phenomenological ansatz incorporating as many symmetries as possible, and
adjust the parameters of these functionals to ground state properties of
characteristic nuclei all over the periodic table (for a recent review see
Ref.~\cite{BHR.03}).

Of particular interest are covariant density functionals \cite{Rin.96,VALR.05}
because they are based on Lorentz invariance. The inclusion of this symmetry
not only allows for the description of the spin-orbit part of the nuclear
interaction in a natural and consistent way, but it also puts considerable
restrictions on the number of parameters in the corresponding functionals, all
without reducing the quality of the agreement with experimental data. A very
successful example is the Relativistic Hartree-Bogoliubov model
\cite{GEL.96,VALR.05}, which combines a density dependence through a
non-linear coupling between the meson fields \cite{BB.77} with pairing
correlations based on an effective interaction of finite range \cite{GEL.96}.

Excited states are described within this formalism by time-dependent density
functional theory \cite{VBR.95}. In the small amplitude limit one obtains the
relativistic Random Phase Approximation (RRPA) \cite{RS.80,RMG.01}. This
method provides a natural framework to investigate collective and
non-collective excitations of $ph$-character. Although several RRPA
implementations have been available since the eighties~\cite{Fur.85}, only
very recently RRPA-based calculations have reached a level on which a
quantitative comparison with experimental data became possible~\cite{MWG.02}.
And even though the self-consistent relativistic mean-field (RMF) framework
has been employed in many studies of deformed nuclei \cite{PRB.87,GRT.90},
applications of the relativistic RRPA method have so far restricted to
spherical nuclei. This is also true for non-relativistic density functionals,
where most of the RPA-calculations are restricted to spherical nuclei. Only
very few deformed RPA calculations based on Skyrme \cite{HHR.98,NKK.06} or
Gogny forces \cite{PGB.07} are available so far.

On the other hand, it is well known that only semi-magic nuclei have spherical
shape, and that most of the other nuclei in the nuclear chart are deformed.
Thus, the description of the collective response of these nuclei can only be
accomplished within a framework where deformation is explicitly taken into
account. From the point of view of nuclear structure the motivation for
deformed RPA calculations is evident. In addition, the nuclear electric dipole
response obtained in this framework provides valuable input for the
calculation of important astrophysical processes \cite{Gor.98}, such as the
$r$- or the $s$-process, that pass though large areas of deformed nuclei.

In this manuscript we report on the extension of relativistic RPA theory to
axially deformed nuclei and its application to the study of collective
excitations. In section~\ref{sec:DFT} we discuss the underlying density
functional and the derivation of the relativistic RPA equations.
Section~\ref{sec:def} deals with specific aspects of RMF and RPA theory in
deformed systems and the evaluation of the relativistic RPA matrix elements in
the basis of axially deformed Dirac spinors. Section \ref{sec:multipole} is
devoted to strength functions and sum rules and in section
\ref{sec:transition} we discuss transition densities in the intrinsic and in
the laboratory frame. Violations of symmetries and the corresponding Goldstone
modes are treated in Section \ref{sec:spurious} and in section
\ref{sec:applications} we show illustrative applications in $^{20}$Ne, in
particular its magnetic and electric dipole response. Finally section
\ref{sec:conclusions} contains the summary and an outlook.

\section{The covariant energy density functional and the relativistic RPA
equations}

\label{sec:DFT}

At the moment the most successful density functionals in nuclear physics are
purely phenomenological. Considering from the beginning as many symmetries as
possible, one starts with a relatively simple ansatz for the energy density
functional \cite{VB.72,DG.80,Wal.74}, which contains a certain number of
phenomenological parameters. One then adjusts these parameters to bulk
properties of nuclear matter and to ground state properties of a few selected
finite nuclei with spherical shape. These sets are then used over the entire
nuclear chart. It turns out that for a good description of the experiment
data, it is crucial to allow for a density dependence in this ansatz. the
concept of density dependence has its origin in more microscopic theories of
the nuclear many-body system, such as Brueckner theory \cite{Bru.54}, which
leads to a density dependent effective interaction in the nuclear interior. In
relativistic models this density dependence was taken into account in the form
of a non-linear meson coupling in Ref.~\cite{BB.77} or in the form of density
dependent meson-nucleon couplings in Ref.~\cite{BT.92}. If the ansatz is
chosen properly and if the adjustment of the phenomenological parameters is
carefully done, the quantitative agreement with available experimental data is
remarkable~\cite{BHR.03,VALR.05}.

In this work we concentrate on relativistic density functional theory
\cite{Rin.96,VALR.05}. These functionals are based on Lorentz invariance. The
basic degrees of freedom are the nucleons described by point-like Dirac
spinors. In order to be consistent with Lorentz invariance and causality, one
has two possibilities for introducing an interaction between these particles.
Either one restricts the theory to zero range interactions, as it is done in
Nambu Jona-Lasinio models ~\cite{NJL.61a}, or one allows for the exchange of
effective mesons. Since it is well know since the early days of Skyrme theory
that pure $\delta$-forces are not sufficient to describe at the same time
nuclear binding energies and radii, and since gradient terms in the Lagrangian
can lead to certain difficulties in the relativistic formulation, historically
the second method was the first to be used \cite{Wal.74,BB.77}. Only recently
relativistic point coupling models with density dependent coupling constants
have been employed successfully in nuclear physics \cite{BMM.02}.

For simplicity we concentrate in this work on meson exchange models with
non-linear meson couplings. Of course, the corresponding equations can be
easily extended to meson coupling models with density dependent vertices
\cite{TW.99,DD-ME1,DD-ME2} or to relativistic point coupling models
\cite{BMM.02}.

In covariant density functionals with meson exchange, the nucleons are
described by Dirac spinors coupled by the exchange of mesons and by the
electromagnetic field through an effective Lagrangian. The starting point for
a phenomenological ansatz is therefore the Walecka model \cite{Wal.74}. The
mesons are classified by there quantum numbers, spin, parity and isospin
($I^{\pi},T$). In the isoscalar channel one has the scalar $\sigma$-meson
($I^{\pi}=0^{+},T=0$), and the vector $\omega$-meson ($I^{\pi}=1^{-},T=0$),
and in the isovector channel one considers only the vector $\rho$-meson
($I^{\pi}=1^{-},T=1$). The $\delta$-meson ($I^{\pi}=0^{+},T=1$) is not
included because, so far, there is not enough data in low energy nuclear
structure physics to fix its parameters uniquely. In addition, the pion is not
taken into account because, again for the sake of simplicity, we work only at
the Hartree level, which forbids the appearance of the parity violating
pion-field. The essential contributions of pionic degrees of freedom by
two-pion exchange are taken care of in a phenomenological way by the $\sigma
$-meson. Therefore, the starting point is an effective Lagrangian density of
the form
\begin{equation}
\mathcal{L}=\mathcal{L}_{N}+\mathcal{L}_{m}+\mathcal{L}_{int}. \label{lagdens}%
\end{equation}
$\mathcal{L}_{N}$ refers to the Lagrangian of the free nucleon
\begin{equation}
\mathcal{L}_{N}=\bar{\psi}(i\gamma^{\mu}\partial_{\mu}-m)\psi,
\end{equation}
where $m$ is the bare nucleon mass and $\psi$ denotes the Dirac spinor.
$\mathcal{L}_{m}$ is the Lagrangian of the free meson fields and the
electromagnetic field
\begin{align}
\mathcal{L}_{m}  &  =\frac{1}{2}\partial_{\mu}\sigma\partial^{\mu}\sigma
-\frac{1}{2}m_{\sigma}^{2}\sigma^{2}-\frac{1}{4}\Omega_{\mu\nu}\Omega^{\mu\nu
}+\frac{1}{2}m_{\omega}^{2}\omega_{\mu}\omega^{\mu}\nonumber\\
&  -\frac{1}{4}\vec{R}_{\mu\nu}\vec{R}^{\mu\nu}+\frac{1}{2}m_{\rho}^{2}%
\vec{\rho}_{\mu}\vec{\rho}^{\mu}-\frac{1}{4}F_{\mu\nu}F^{\mu\nu}%
\end{align}
with the corresponding masses $m_{\sigma}$, $m_{\omega}$, $m_{\rho}$, and the
field tensors
\begin{align}
\Omega_{\mu\nu}  &  =\partial_{\mu}\omega_{\nu}-\partial_{\nu}\omega_{\mu
},\nonumber\\
\vec{R}_{\mu\nu}  &  =\partial_{\mu}\vec{\rho}_{\nu}-\partial_{\nu}\vec{\rho
}_{\mu},\\
F_{\mu\nu}  &  =\partial_{\mu}A_{\nu}-\partial_{\nu}A_{\mu},\nonumber
\end{align}
where arrows denote isovectors and boldface symbols vectors in 3-dimensional
$r$-space. The interaction Lagrangian $\mathcal{L}_{int}$ is given by minimal
coupling terms
\begin{equation}
\mathcal{L}_{int}=-g_{\sigma}\bar{\psi}\Gamma_{\sigma}\sigma\psi-g_{\omega
}\bar{\psi}\Gamma_{\omega}^{\mu}\omega_{\mu}\psi-g_{\rho}\bar{\psi}\vec
{\Gamma}_{\rho}^{\mu}\vec{\rho}_{\mu}\psi-e\bar{\psi}\Gamma_{e}^{\mu}A_{\mu
}\psi
\end{equation}
with the vertices
\begin{equation}
\Gamma_{\sigma}=1,\quad\Gamma_{\omega}^{\mu}=\gamma^{\mu},\quad\vec{\Gamma
}_{\rho}^{\mu}=\gamma^{\mu}\vec{\tau},\quad\Gamma_{e}^{\mu}=\dfrac{1}%
{2}(1-\tau_{3})\gamma^{\mu}, \label{rmf:Lvertex}%
\end{equation}
where $g_{\sigma}$, $g_{\omega}$, $g_{\rho}$ and $e$ are the respective
coupling constants for the $\sigma$, $\omega$, $\vec{\rho}$ and photon fields.
This yields to
\begin{equation}
\mathcal{L}_{int}=-\sum_{m}g_{m}\bar{\psi}\Gamma_{m}\phi_{m}\psi,
\end{equation}
where the index $m$ runs over the various meson and electromagnetic fields,
and also over the Lorentz index for vector mesons and isospin indices for
mesons carrying isospin
\begin{equation}
\phi_{m}=(\sigma,\omega^{\mu},\vec{\rho}^{\mu},A^{\mu}),\qquad\Gamma
_{m}=(\Gamma_{\sigma},\Gamma_{\omega}^{\mu},\vec{\Gamma}_{\rho}^{\mu}%
,\Gamma_{e}^{\mu}).
\end{equation}
Already in the earliest applications of the RMF framework it was realized,
however, that this simple linear interaction density functional did not
provide a quantitative description of complex nuclear systems; an effective
density dependence needs to be introduced. Historically, the first
\cite{BB.77} was the inclusion of non-linear self-interaction terms in the
meson part of the Lagrangian in the form of a quartic $\sigma$ potential
\begin{equation}
\frac{1}{2}m_{\sigma}^{2}\sigma^{2}+U(\sigma)
\end{equation}
with
\begin{equation}
U(\sigma)=\frac{g_{2}}{3}\sigma^{3}+\frac{g_{3}}{4}\sigma^{4},
\label{rmf:nonlinearU}%
\end{equation}
which includes the non-linear $\sigma$ self-interactions with two additional
parameters $g_{2}$ and $g_{3}$. This particular form of the non-linear
potential has become standard in applications of RMF functionals, although
additional non-linear interaction terms, both in the isoscalar and isovector
channels, have been considered over the years \cite{Bod.91,TM1,SFM.00,HP.01}.

Two other approaches, of more recent development, can also be found in the
literature, based on the introduction of the density dependence directly in
the coupling constants \cite{TW.99,DD-ME1,DD-ME2} and on the expansion of the
meson propagators into zero-range couplings and gradient corrections terms
\cite{BMM.02}. For the sake of simplicity we will restrict the discussion in
this work to non-linear density functionals, always taking the NL3 \cite{NL3}
parameter set as the force of choice. Also, the explicit inclusion of the
non-linear meson potential $U(\sigma)$ is generally avoided in order to keep
the formulation as clean as possible. But, of course, it is included in all
numerical calculations of results presented in the manuscript, and it shall be
explicitly mentioned at certain important points in the following discussions.

The Hamiltonian density can be derived from the Lagrangian density of
Eq.~\eqref{lagdens} as the (0,0) component of the energy-momentum tensor
\begin{equation}
\mathcal{H}=T^{00}=\frac{\partial\mathcal{L}}{\partial\dot{q}_{j}}\dot{q}%
_{j}-\mathcal{L},
\end{equation}
leading the to the energy functional
\begin{equation}
E[\hat{\rho},\phi]=\int\mathcal{H}d^{3}r.
\end{equation}
Following the Kohn-Sham approach \cite{KS.65,KS.65a}, one can express the
relativistic energy density $E$ as a functional of the relativistic single
particle density matrix
\begin{equation}
\hat{\rho}(\boldsymbol{r},\boldsymbol{r}^{\prime},t)=%
{\displaystyle\sum\limits_{i}^{A}}
\psi_{i}(\boldsymbol{r}\mathbf{,}t)\psi_{i}^{\dagger}(\boldsymbol{r}^{\prime
},t) \label{rmf:rho}%
\end{equation}
and the meson fields $\phi=(\sigma$,$\omega$,$\rho$,$\gamma)$. The sum over
$i$ in Eq.~(\ref{rmf:rho}) runs over the orbits in the Dirac sea
(\textit{no-sea approximation}, see below). Considering the 4-dimensional
Dirac spinor $\psi_{i}$ as a column vector and $\psi_{i}^{\dagger}$ as a row
vector one concludes that $\hat{\rho}(\boldsymbol{r},\boldsymbol{r}^{\prime
},t)$ is a 4x4 matrix in Dirac space. This leads to the standard relativistic
energy density functional
\begin{align}
E_{RMF}[\hat{\rho},\phi]  &  =\mathrm{Tr}[(-i\boldsymbol{\alpha}%
\boldsymbol{\nabla}+\beta m)\hat{\rho}]+\mathrm{Tr}[(\beta\Gamma_{m}\phi
_{m})\hat{\rho}]\nonumber\\
&  \pm\frac{1}{2}\int d^{3}r\left[  (\partial_{\mu}\phi_{m})^{2}+m_{m}%
^{2}\right]  , \label{rmf:energyfunctional}%
\end{align}
where summation over the different mesons is implied, and the trace operation
involves summation over Dirac indices and an integral over the whole space.
$\Gamma_{m}$ describes the structure of the meson-nucleon interaction. The
upper sign in Eq.~(\ref{rmf:energyfunctional}) holds for the scalar mesons and
the lower sign for the vector mesons. At the mean field level the mesons are
treated as classical fields. The nucleons, described by a Slater determinant
$|\Phi\rangle$ of single-particle wave functions, move independently in these
classical meson fields. One can thus apply the classical time-dependent
variational principle
\begin{equation}
\delta\int_{t_{1}}^{t_{2}}dt\{\langle\Phi|i\partial_{t}|\Phi\rangle
-E[\hat{\rho},\phi]\}=0,
\end{equation}
which leads to the equations of motion
\begin{align}
i\partial_{t}\hat{\rho}  &  =\left[  h[\hat{\rho},\phi],\hat{\rho}\right]
,\label{rmf:feqnmotion}\\
\left[  \partial^{\nu}\partial_{\nu}+m_{m}^{2}\right]  \phi_{m}  &
=\mp\mathrm{Tr}\left[  \beta\Gamma_{m}\hat{\rho}\right]  ,
\label{rmf:beqnmotion}%
\end{align}
where the single particle effective Dirac Hamiltonian $\hat{h}$ is the
functional derivative of the energy with respect to the single particle
density
\begin{equation}
\hat{h}[\hat{\rho},\phi]=\frac{\delta E[\hat{\rho},\phi]}{\delta\hat{\rho}%
}=(-i\boldsymbol{\alpha}\boldsymbol{\nabla}+\beta m)+%
{\displaystyle\sum\limits_{m}}
\beta\Gamma_{m}\phi_{m}. \label{rmf:dirach}%
\end{equation}
The time-dependent Dirac equation for the nucleons reads
\begin{equation}
\left[  \gamma^{\mu}\left(  i\partial_{\mu}+V_{\mu}\right)  +m+S\right]
\psi_{k}=0 \label{tdde}%
\end{equation}
with the scalar $S$ and vector $V_{\mu}$ potentials
\begin{align}
S(\boldsymbol{r},t)  &  =g_{\sigma}\sigma(\boldsymbol{r},t),\\
V_{\mu}(\boldsymbol{r},t)  &  =g_{\omega}\omega_{\mu}(\boldsymbol{r}%
,t)+g_{\rho}\vec{\tau}\vec{\rho}_{\mu}(\boldsymbol{r},t)+eA_{\mu
}(\boldsymbol{r},t)\frac{1-\tau_{3}}{2}. \nonumber\label{tdme}%
\end{align}
and the time-dependent meson equations have the form%
\begin{align}
\left[  \partial^{\nu}\partial_{\nu}+m_{\sigma}^{2}\right]  \sigma &
=-g_{\sigma}\rho_{s}\\
\left[  \partial^{\nu}\partial_{\nu}+m_{\omega}^{2}\right]  \omega^{\mu}  &
=+g_{\omega}j^{\mu}\\
\left[  \partial^{\nu}\partial_{\nu}+m_{\rho}^{2}\right]  \vec{\rho}^{\mu}  &
=+g_{\rho}\vec{j}_{\text{tv}}^{\mu}\\
\left[  \partial^{\nu}\partial_{\nu}\right]  A^{\mu}  &  =+ej_{\text{c}}^{\mu}%
\end{align}
with the sources%
\begin{align}
&  \mathrm{scalar-isoscalar}\qquad\rho_{s}=\sum_{i}^{A}\bar{\psi}_{i}\psi
_{i},\\
&  \mathrm{vector-isoscalar}\qquad j^{\mu}=\sum_{i}^{A}\bar{\psi}_{i}%
\gamma^{\mu}\psi_{i},\\
&  \mathrm{vector-isovector}\qquad\vec{j}_{\text{tv}}^{\mu}=\sum_{i}^{A}%
\bar{\psi}_{i}\gamma^{\mu}\vec{\tau}\psi_{i},\\
&  \mathrm{electromagnetic}\qquad j_{\text{c}}^{\mu}=\sum_{i}^{A}\bar{\psi
}_{i}\gamma^{\mu}\frac{1}{2}(1-\tau_{3})\psi_{i}%
\end{align}
In order to describe the ground state properties of even-even nuclei, one has
to look for stationary time-reversal invariant solutions of the equations of
motion Eq.~\eqref{rmf:feqnmotion} and Eq.~\eqref{rmf:beqnmotion}. The nucleon
wave functions are then the eigenvectors of the stationary Dirac equation,
\begin{equation}
\left[  -i\boldsymbol{\alpha}\boldsymbol{\nabla}+V_{0}+\beta(m+S)\right]
\psi_{k}=\varepsilon_{k}\psi_{k} \label{rmf:diraceqn}%
\end{equation}
which yields the single particle energies $\varepsilon_{k}$ as eigenvalues.

The meson fields and the Coulomb potential obey the Helmholtz and Laplace
equations%
\begin{align}
\left[  -\Delta+m_{\sigma}^{2}\right]  \sigma &  =-g_{s}\rho_{s}%
,\label{rmf:mesoneqnstat1}\\
\left[  -\Delta+m_{\omega}^{2}\right]  \omega^{0}  &  =+g_{\omega}%
\rho_{\text{v}},\\
\left[  -\Delta+m_{\rho}^{2}\right]  \rho^{0}  &  =+g_{\rho}\rho_{\text{tv}%
},\\
\left[  -\Delta\right]  A^{0}  &  =+e\rho_{\text{c}},
\label{rmf:mesoneqnstat4}%
\end{align}
with the following the source densities
\begin{align}
&  \mathrm{scalar-isoscalar}\qquad\rho_{s}=\sum_{i}^{A}\bar{\psi}_{i}\psi
_{i},\\
&  \mathrm{vector-isoscalar}\qquad\rho_{\text{v}}=\sum_{i}^{A}\psi
_{i}^{\dagger}\psi_{i},\\
&  \mathrm{vector-isovector}\qquad\rho_{\text{tv}}=\sum_{i}^{A}\psi
_{i}^{\dagger}\tau_{3}\psi_{i},\\
&  \mathrm{electromagnetic}\qquad\rho_{\text{c}}=\sum_{i}^{A}\psi_{i}%
^{\dagger}\frac{1}{2}(1-\tau_{3})\psi_{i} \label{rmf:densities}%
\end{align}
Eq.~\eqref{rmf:diraceqn} together with
Eqs.~\eqref{rmf:mesoneqnstat1}-\eqref{rmf:mesoneqnstat4} pose a
self-consistent problem which is readily solved by iteration. With the
resulting density $\hat{\rho}$ and fields $\phi$, the total energy of the
system can be calculated using Eq.~\eqref{rmf:energyfunctional}. Radii and
other bulk properties of the nucleus can be derived as well.

An important point of the present versions of covariant density functional
theory is the \textit{no-sea approximation}, i.e. in the calculation of the
sources for the meson equations
\eqref{rmf:mesoneqnstat1}-\eqref{rmf:mesoneqnstat4} only positive energy
spinors are included in the summation. In a fully relativistic description
also the negative energy states from the Dirac sea would have to be included.
However, this would lead to divergent terms which have to be treated by a
proper renormalization procedure in nuclear matter \cite{CW.74,Chin.77} or in
finite nuclei \cite{HS.84,BS.84,Per.87,HEM.00}. Numerical studies have shown
that effects due to vacuum polarization can be as large as 20\%-30\%. Their
inclusion requires a readjustment of the parameter set for the effective
Lagrangian that leads to approximately the same results as if they were
neglected \cite{HS.84,Was.88,ZMR.91}. This means that in a phenomenological
theory based on the \textit{no-sea approximation}, where the parameters are
adjusted to experimental data, vacuum polarization is not neglected, it is
just taken into account in the phenomenological parameters in a global fashion
and the \textit{no-sea approximation} is in reality not an approximation. It
is used in all successful applications of covariant DFT. This has, however,
serious consequences for the calculation of excited states in the RPA
\cite{RMG.01}.

The vibrational response of the system can be studied considering harmonic
oscillations with small amplitude and with eigen-frequencies $\Omega_{\nu}$
around the stationary ground $\hat{\rho}^{(0)}$. In this case, the
time-dependent density can be written as
\begin{equation}
\hat{\rho}(t)=\hat{\rho}^{(0)}+(\delta\hat{\rho}^{(\nu)}e^{-i\Omega_{\nu}%
t}+h.c.).
\end{equation}
Imposing the condition that $\hat{\rho}$ is a projector at all times, the
transition density matrices $\delta\hat{\rho}^{(\nu)}$ have only matrix
elements which connect occupied and unoccupied states \cite{RS.80}
\begin{align}
X_{mi}^{(\nu)}  &  =\delta\rho_{mi}^{(\nu)}=\langle0|a_{i}^{\dagger}a_{m}%
|\nu\rangle\nonumber\\
Y_{mi}^{(\nu)}  &  =\delta\rho_{im}^{(\nu)}=\langle0|a_{m}^{\dagger}a_{i}%
|\nu\rangle
\end{align}
with respect to the stationary solution $\hat{\rho}^{(0)}.$ In the
non-relativistic case these are only $ph$- and $hp$- matrix elements, i.e.,
the index $i$ $\ $runs over all levels in the Fermi sea and the index $m$ runs
over all empty levels above the Fermi sea.

In linear order, the equations of motion \eqref{rmf:beqnmotion} can be written
down as the RPA equations in their standard matrix form
\begin{equation}
\left(
\begin{array}
[c]{cc}%
A & B\\
-B^{\ast} & -A^{\ast}%
\end{array}
\right)  \left(
\begin{array}
[c]{c}%
X^{(\nu)}\\
Y^{(\nu)}%
\end{array}
\right)  =\Omega^{(\nu)}\left(
\begin{array}
[c]{c}%
X^{(\nu)}\\
Y^{(\nu)}%
\end{array}
\right)  , \label{rpa:eqn}%
\end{equation}
where the $X^{(\nu)}$ refers to the forward amplitude transition density and
$Y^{(\nu)}$ to the backward amplitude. The forward amplitude is thus
associated with the creation and the backwards amplitude with the destruction
of a $ph$-pair.

In the relativistic case the situation is more complicated. Because of the
\textit{no-sea approximation} in the RMF-model, the Dirac sea is empty.
Therefore, we have to consider in relativistic RPA not only the $ph$- (and
$hp$-) matrix elements of $\delta\hat{\rho}$, but also the matrix elements
$\delta\hat{\rho}_{ah}$ and $\delta\hat{\rho}_{ha}$ connecting states in the
Dirac sea with those in the Fermi sea. The index $i\ $ in the amplitudes
$X_{mi}^{(\nu)}$ and $Y_{mi}^{(\nu)}$, and therefore in the RPA equation
(\ref{rpa:eqn}), runs again over all the levels in the Fermi sea; however, the
index $m$ runs now over all the levels above the Fermi sea and over all the
levels in the Dirac sea. This means we have not only to take particles in the
Fermi sea and put them in the empty levels above the Fermi surface, but we
have to consider also configurations where we form holes in the Fermi see and
occupy empty levels in the Dirac sea. At a first glance this seems to be
completely unphysical, because according to Dirac, the Dirac sea should be
filled with particles. It turns out, however, that this is not the case.
Considering the time-dependent RMF equations, the Dirac sea depends on time
and the no-sea approximation should be realized at every point in time. In
fact, solving these equations we consider only the time-evolution of the
levels $\psi_{i}(t)$ in the Fermi sea. The corresponding time-dependent levels
in the Dirac sea stay empty for all times \cite{RMG.01}. When we describe this
situation in the static basis, it is a mathematical consequence of the
completeness of the basis, that one has to include also the $ah$%
-configurations. If one neglects those configurations, self-consistency is
violated and one does not preserve the nice properties of RPA, such as current
conservation \cite{DF.90} and the separation of the Goldstone modes (spurious
states) from the other physical solutions.

Neglecting the $ah$-configurations in the RPA equations also leads in specific
cases to highly unphysical results, as for instance shifts in the energy of
the GMR in $^{208}$Pb from the experimental value at 14 MeV down to 2-3 MeV
\cite{MGT.97}.

Taking into account also $ah$-configurations renders the solution of the
relativistic RPA equations much more complicated than in the non-relativistic
case, because (i) the dimension of these equations increases considerably and
(ii) the matrix $A\pm B$ is no longer positive definite and therefore the
non-Hermitian matrix diagonalization problem can not be transformed into a
Hermitian problem of half dimension, as discussed for instance in
Ref.~\cite{RS.80}. Only recently it has been shown that the relativistic
RPA-equations can be reduced to a non-Hermitian diagonalization problem of
half dimension \cite{Pap.07}.

For two different RPA excited states $\nu$ and $\nu^{\prime}$, the following
orthogonality relation holds
\begin{equation}
\sum_{mi}X_{mi}^{(\nu)\ast}X_{mi}^{(\nu^{\prime})}-Y_{mi}^{(\nu)\ast}%
Y_{mi}^{(\nu^{\prime})}=\delta_{\nu\nu^{\prime}},\label{rpa:rpanorm}%
\end{equation}
which can be used to normalize the eigen vectors $(X^{(\nu)},Y^{(\nu)})$.
Within the RPA-approximation the transition matrix elements for a one body
operator $\hat{\mathcal{O}}$ between the excited state $|\nu\rangle$ and the
ground state $|0\rangle$ are given by
\begin{equation}
\langle0|\hat{\mathcal{O}}|\nu\rangle=%
{\displaystyle\sum\limits_{mi}}
\mathcal{O}_{mi}X_{mi}^{(\nu)}+\mathcal{O}_{mi}^{\ast}Y_{mi}^{(\nu
)}.\label{rpa:obo}%
\end{equation}
The RPA matrices $A$ and $B$ read
\begin{align}
A_{mi,nj} &  =(\varepsilon_{m}-\varepsilon_{i})\delta_{mn}\delta_{ij}%
+V_{mjin}^{ph},\label{rpa:RPA}\\
B_{mi,nj} &  =V_{mnij}^{ph},\label{rpa:RPA-B}%
\end{align}
where the matrix elements $V_{kl^{\prime}k^{\prime}l}^{ph}$ are the second
derivatives of the energy functional with respect to the single particle
density
\begin{equation}
V_{kl^{\prime}k^{\prime}l}^{ph}=\langle kl^{\prime}|\hat{V}^{ph}|k^{\prime
}l\rangle=\frac{\delta^{2}E}{\delta\rho_{k^{\prime}k}\delta\rho_{ll^{\prime}}%
}\label{rpa:vphph}%
\end{equation}
and $\hat{V}^{ph}$ is the \textit{effective interaction}. As we have seen, the
mean-field ground state is characterized by the stationary density matrix
$\hat{\rho}^{(0)}$ and by the meson fields $\phi^{(0)}$, which, up to this
point, have been treated as independent variables connected to the density by
the equations of motion in Eq.~\eqref{rmf:beqnmotion}.

In order to describe small oscillations self-consistently it turns out to be
useful to eliminate the meson degrees of freedom from the energy functional,
such that the fermion equation of motion (approximated by the RPA equation
(\ref{rpa:RPA})) is closed, i.e., the residual interaction has to be expressed
as a functional of the generalized density $\hat{\rho}$ only. This elimination
of the meson degrees of freedom is possible only in the limit of small
amplitudes,
\begin{align}
\phi &  =\phi^{(0)}+\delta\phi,\nonumber\\
\hat{\rho}  &  =\hat{\rho}^{(0)}+\delta\hat{\rho}, \label{smallampl}%
\end{align}
where $\delta\hat{\rho}$ and $\delta\phi$ are small deviations from the ground
state values $\hat{\rho}^{(0)}$ and $\phi^{(0)}$. Substituting this expansion
in the Klein-Gordon equations \eqref{rmf:beqnmotion} and retaining only the
first order in $\delta\hat{\rho}$ we find
\begin{equation}
\left[  \partial_{\mu}\partial^{\mu}+m_{m}^{2})\right]  \delta\phi_{m}=\mp
g_{m}\delta\rho_{m},
\end{equation}
with the local densities, the sources for the various meson fields are given
by $\delta\rho_{m}(\mathbf{r})=(\delta\rho_{\text{s}}(\mathbf{r)},\delta
\rho_{\text{v}}(\mathbf{r),}\delta\rho_{\text{vt}}(\mathbf{r),}\delta
\rho_{\text{c}}(\mathbf{r)})$ for $m=(\sigma,\omega,\rho,A)$. Neglecting
retardation effects (i.e. neglecting $\partial_{t}^{2}$) one finds for the
linearized equations of motion for the mesons
\begin{equation}
\left[  -\Delta+m_{m}^{2})\right]  \delta\phi_{m}=\mp g_{m}\delta\rho_{m}.
\label{rpa:linmeseq}%
\end{equation}
This approximation is meaningful only at small energies, as compared to the
meson masses, where the short range of the corresponding meson exchange forces
guarantees that retardation effects can be neglected. A formal solution for
\eqref{rpa:linmeseq} can be written as
\begin{equation}
\delta\phi_{m}(\boldsymbol{r})=\mp\int d^{3}r^{\prime}g_{m}G_{m}%
(\boldsymbol{r},\boldsymbol{r}^{\prime})\delta\rho_{m}(\boldsymbol{r}^{\prime
})
\end{equation}
which allows us to decompose the residual interaction $\hat{V}^{ph}$ in
various meson exchange forces
\begin{equation}
\hat{V}^{ph}=\hat{V}_{\sigma}+\hat{V}_{\omega}+\hat{V}_{\rho}+\hat{V}_{\gamma}
\label{rpa:vintdecomp}%
\end{equation}
with
\begin{equation}
\hat{V}_{m}^{{}}(1,2)=\mp g_{m}^{2}(\beta\Gamma_{m}^{{}})^{(1)}G_{m}%
(\boldsymbol{r}_{1},\boldsymbol{r}_{2})(\beta\Gamma_{m}^{{}})^{(2)}.
\label{rpa:E51}%
\end{equation}
For linear meson couplings the propagator $G_{m}$ obeys the Helmholtz
equation
\begin{equation}
(-\Delta+m_{m}^{2})G_{m}(\boldsymbol{r},\boldsymbol{r}^{\prime})=\delta
(\boldsymbol{r}-\boldsymbol{r}^{\prime}), \label{rpa:propaeq}%
\end{equation}
and has Yukawa form
\begin{equation}
G_{m}(\boldsymbol{r}_{1},\boldsymbol{r}_{2})=\frac{1}{4\pi}\frac
{e^{-m_{m}|\boldsymbol{r}_{1}-\boldsymbol{r}_{2}|}}{|\boldsymbol{r}%
_{1}-\boldsymbol{r}_{2}|}.
\end{equation}
The vertices $\beta\Gamma_{m}$ reflect the different covariant structures of
the fields as defined in Eq.~\eqref{rmf:Lvertex}. Combining the spatial
coordinates $\boldsymbol{r}$ and the Dirac index $\alpha=1\dots4$ to the
coordinate $1=(\boldsymbol{r}_{1},\alpha_{1})$ we can express the relativistic
two-body interactions in the following way

\begin{itemize}
\item for the $\sigma$-exchange
\begin{equation}
\hat{V}_{\sigma}(1,2)=-\frac{g_{\sigma}^{2}}{4\pi}\beta^{(1)}\beta^{(2)}%
\frac{e^{-m_{m}|\boldsymbol{r}_{1}-\boldsymbol{r}_{2}|}}{|\boldsymbol{r}%
_{1}-\boldsymbol{r}_{2}|},
\end{equation}

\item for the $\omega$-exchange
\begin{equation}
\hat{V}_{\omega}(1,2)=\frac{g_{\omega}^{2}}{4\pi}(1-\boldsymbol{\alpha}%
^{(1)}\boldsymbol{\alpha}^{(2)})\frac{e^{-m_{m}|\boldsymbol{r}_{1}%
-\boldsymbol{r}_{2}|}}{|\boldsymbol{r}_{1}-\boldsymbol{r}_{2}|},
\end{equation}

\item for the $\rho$-exchange
\begin{equation}
\hat{V}_{\rho}(1,2)=\frac{g_{\rho}^{2}}{4\pi}\vec{\tau}^{(1)}\vec{\tau}%
^{(2)}(1-\boldsymbol{\alpha}^{(1)}\boldsymbol{\alpha}^{(2)})\frac
{e^{-m_{m}|\boldsymbol{r}_{1}-\boldsymbol{r}_{2}|}}{|\boldsymbol{r}%
_{1}-\boldsymbol{r}_{2}|},
\end{equation}

\item for the electromagnetic interaction
\begin{equation}
\hat{V}_{\text{em}}(1,2)=\frac{e^{2}}{4\pi}\frac{1-\tau_{3}^{(1)}}{2}%
\frac{1-\tau_{3}^{(2)}}{2}\frac{1-\boldsymbol{\alpha}^{(1)}\boldsymbol{\alpha
}^{(2)}}{|\boldsymbol{r}_{1}-\boldsymbol{r}_{2}|}%
\end{equation}

\end{itemize}

In the case of non-linear meson couplings the Klein-Gordon equation
\ref{rpa:linmeseq} is replaced by%
\begin{equation}
\left[  -\Delta+m_{\sigma}^{2}\right]  \sigma+U^{\prime}(\sigma)=-g_{s}%
\rho_{s}.
\end{equation}
Considering small oscillations around the static solution $\sigma^{(0)}$, it
leads, instead of Eq.~\eqref{rpa:linmeseq}, to
\begin{equation}
\left[  -\Delta+m_{\sigma}^{2}+W(\boldsymbol{r})\right]  \delta\sigma
=-g_{\sigma}\delta\rho_{\text{s}},
\end{equation}
with%
\begin{equation}
W(\boldsymbol{r})=U^{\prime\prime}(\sigma^{(0)}(\boldsymbol{r}))
\end{equation}
and the propagator $G_{\sigma}(\boldsymbol{r},\boldsymbol{r}^{\prime})$ obeys
the equation
\begin{equation}
\left[  -\Delta+m_{\sigma}^{2}+W(\boldsymbol{r})\right]  G_{\sigma
}(\boldsymbol{r},\boldsymbol{r}^{\prime})=\delta(\boldsymbol{r}-\boldsymbol{r}%
^{\prime})\label{rpa:nlpropaeq}%
\end{equation}
which cannot be solved analytically. More details, how to determine this
propagator numerically are given in Appendix \ref{app:B}.

\section{RMF+RRPA in deformed nuclei}

\label{sec:def}

The fact that nuclei can be deformed was already emphasized by Niels Bohr in
his classic paper on the nuclear liquid-drop model \cite{Boh.36}, where he
introduced the concept of nuclear shape vibrations. If a system is deformed,
its spatial density is anisotropic, so it is possible to define its
orientation as a whole, and this naturally leads to the presence of collective
rotational modes. In 1950, Rainwater \cite{Rai.50} observed that the
experimentally measured large quadrupole moments of nuclei could be explained
in terms of the deformed shell model, i.e., the extension of the spherical
shell model to the case of a deformed average potential. In a following paper
\cite{Boh.51}, Age Bohr formulated the basis of the particle-rotor model, and
introduced the concept of an intrinsic (body-fixed) nuclear system defined by
means of shape deformations regarding nuclear shape and orientation as
dynamical variables. The basic microscopic mechanism leading to the existence
of nuclear deformations was proposed by A. Bohr \cite{Boh.52}, stating that
the strong coupling of nuclear surface oscillations to the motion of
individual nucleons is the reason for the observed static deformations in
nuclei. Nowadays, the deformation mechanism in nuclei is well understood
\cite{RS.80}: for sufficiently high level density in the vicinity of the Fermi
surface, or for sufficiently strong residual interaction, the first $2^{+}$
excited state (a quadrupole surface phonon) is shifted down to zero energy (it
\textquotedblleft freezes out\textquotedblright), effectively creating a
condensate of quadrupole phonons and such giving rise to a static deformation
of the mean-field ground state.

In order to calculate excitations in deformed nuclei RPA theory outlined in
the previous section can be used. It is important to remember, however, that
these excitations are intrinsic in as much as they are relative to the local
deformed ground-state. Nevertheless, the application of RPA for the
calculation of intrinsic excitations of deformed nuclei is formally completely
analogous to that for spherical nuclei. The only difference is that one has
now single particle orbitals violating rotational symmetry, i.e. having no
good angular momentum. For this reason it is not possible to apply group
theory and to reduce the dimension of the RPA matrix by angular momentum
coupling techniques. Only in the case of axial symmetry, reductions are
possible which using the good quantum number $K$, the projection of the total
angular momentum on the symmetry axis.

The introduction of a deformed intrinsic state in DFT is straightforward. Let
us suppose that there exist a symmetry operator $\mathcal{S}$ such that the
energy density functional is invariant under the symmetry transformations
$e^{i\alpha\mathcal{S}}$, i.e. for a transformed density $\hat{\rho}_{\alpha
}$
\begin{equation}
\hat{\rho}_{\alpha}=e^{-i\alpha\mathcal{S}}\,\hat{\rho}\,e^{i\alpha
\mathcal{S}} \label{def:symtran}%
\end{equation}
we have
\begin{equation}
E[\hat{\rho}_{\alpha}]=E\left[  e^{-i\alpha\mathcal{S}}\,\hat{\rho
}\,e^{i\alpha\mathcal{S}}\right]  =E[\hat{\rho}] \label{def:syminv}%
\end{equation}
Examples of such a symmetry in even-even nuclear systems would be rotational
and translational symmetries and the third component of isospin (i.e. the
charge). If the density has the same symmetry, $\hat{\rho}_{\alpha} =\hat
{\rho}$, we can restrict the set of variational densities to those with this
symmetry. However, such a symmetric solution is not necessarily at the minimum
in the energy surface defined by $E[\hat{\rho}]$, that is, the best solution.
Because of the non-linearity of the variational Eq.~\eqref{rmf:diraceqn} it is
possible that the solution breaks the symmetry spontaneously, i.e., the energy
density is invariant under $\mathcal{S}$-transformations but the density is
not $\hat{\rho}_{\alpha}\neq\rho$.

Rotations are one of such continuous symmetry transformations. Nuclei with
semi-closed shells have pairing correlations, and the solution with lowest
energy of the variational Hartree-Bogoliubov (HB) or Hartree-BCS (HBCS)
equations have a spherical intrinsic density distribution. One can always
write their ground state wave function as a rotationally invariant product
state of HB or BCS type . On the other hand, most nuclei throughout the
periodic table have open shells for both types of particles, and thus due to
the strongly attractive seniority breaking proton-neutron interaction their
respective intrinsic single particle densities are usually not invariant under
rotations. Nevertheless most of the nuclei have minima with axial symmetric
density distributions. There are only few cases with pronounced triaxial deformations.

The present investigation is restricted to nuclei that can be adequately
described by a variational wave function with axial symmetry, and so the
projection of the angular momentum $\Omega$ on the symmetry axis is a
conserved quantity. It is therefore convenient to use cylindrical coordinates%
\begin{equation}
\boldsymbol{r}=(r\cos\varphi,r\sin\varphi,z), \label{cylindrical-r}%
\end{equation}
were, as usual, the symmetry axis is labeled as the $z$-axis. Note, that $r$
is the distance from the symmetry axis, not the distance from the origin. For
reasons of simplicity we avoid in this manuscript, as far as possible, the
notation $r_{\perp}$. The single-particle Dirac spinors $\psi_{k}$, solution
of Eq.~\eqref{rmf:diraceqn}, are then characterized by the angular momentum
projection $\Omega$, the parity $\pi$ and the isospin projection $t$,
\begin{equation}
\psi_{k}(\mathbf{r})=\frac{1}{\sqrt{2\pi}}\left(
\begin{array}
[c]{c}%
f_{k}^{+}(r,z)e^{i(\Omega_{i}-1/2)\phi}\\
f_{k}^{-}(r,z)e^{i(\Omega_{i}+1/2)\phi}\\
ig_{k}^{+}(r,z)e^{i(\Omega_{i}-1/2)\phi}\\
ig_{k}^{-}(r,z)e^{i(\Omega_{i}+1/2)\phi}\\
\end{array}
\right)  \chi_{t_{k}(t)} \label{def:wavef}%
\end{equation}
For even-even nuclei, for each solution $\psi_{k}$ with positive $\Omega_{k}$
there exists a time-reversed one with the same energy that will be denoted by
a bar%
\begin{equation}
\bar{k}:=\{\epsilon_{k},\,-\Omega_{k},\,\pi_{k}\} \label{def:timereversed}%
\end{equation}
The time reversal operator has the usual form $i\gamma^{3}\gamma^{1}\hat{K}$,
where $\hat{K}$ is the complex conjugation.

\subsection{Configuration space for the RPA equation}

The rows and columns of the RPA matrix in Eq.~\eqref{rpa:eqn} are labeled by
all the possible $ph$- and $ah$-pairs that can be formed using the
single-particle spinor solutions of the static problem. Since the total
angular momentum is no longer a good quantum number, we cannot take advantage
of angular momentum techniques when forming these pairs. Only axially symmetry
and parity is left. The full RPA-matrix can be thus reduced to blocks with
good quantum numbers $K$ and $\pi$. In particular, this means that the RPA
matrix elements $V_{mjin}^{ph}$ must obey the following selection rules
\begin{align}
\Omega_{m}-\Omega_{i}  &  =\Omega_{n}-\Omega_{j}=K\label{def:angparcon}\\
\pi_{m}\pi_{i}  &  =\pi_{n}\pi_{j}=\pi\label{def:parity}%
\end{align}
Thus we can define RPA phonon operators
\begin{equation}
Q_{\nu,K^{\pi}}^{+}=\sum_{mi}X_{mi}^{(\nu)}a_{m}^{\dagger}a_{i}^{{}}%
-Y_{mi}^{(\nu)}a_{\bar{\imath}}^{\dagger}a_{\bar{m}}^{{}} \label{def:Kpair}%
\end{equation}
as linear combination of pairs with good angular momentum projection $K$ and
parity $\pi$. This mean that the sum runs only over pairs $mi$ such that the
conditions (\ref{def:angparcon}) and (\ref{def:parity}) are satisfied and the
different excitation modes
\begin{equation}
|\nu,K^{\pi}\rangle=Q_{\nu,K^{\pi}}^{+}|0) \label{def:RPA-wavefunctions}%
\end{equation}
can be labeled by the quantum numbers
\begin{equation}
K^{\pi}=0^{\pm},1^{\pm},2^{\pm},\cdots
\end{equation}
where
\begin{equation}
K^{\pi}=(\Omega_{m}-\Omega_{i})^{(\pi_{m}\pi_{i})} \label{def:pairs}%
\end{equation}
One has to be careful handling time reversal symmetry in the case of coupling
to $K=0$, where for each pair of the form of Eq.~\eqref{def:pairs} there
exists the time reversed one
\begin{equation}
K^{\pi}=(-\Omega_{\bar{m}}+\Omega_{\bar{\imath}})^{(\pi_{m}\pi_{i})}%
=(\Omega_{m}-\Omega_{i})^{(\pi_{m}\pi_{i})}%
\end{equation}
with the same energy that also satisfies Eq.~\eqref{def:angparcon}, and has to
be considered explicitly when calculating the matrix elements.

\subsection{Evaluation of the RPA matrix elements}

As we have seen in Eq.~\eqref{rpa:vphph}, the matrix elements of the residual
interaction can be derived from the energy functional as the second derivative
with respect to the density. In the case of meson exchange models this
interaction in Eqs.~\eqref{rpa:vintdecomp} and \eqref{rpa:E51}. The index $m$ runs
over the various mesons, but, for vector mesons, also over the Minkowski index
$\mu$. For simplicity we therefore use in the following for the vertices only
the Minkowski index $\mu$ instead of $m$. In this case the interaction has the
form
\begin{equation}
\hat{V}_{m}(1,2)=\mp g_{m}^{2}\beta^{(1)}\Gamma_{{}}^{\mu(1)}G_{m}%
(\boldsymbol{r}_{1},\boldsymbol{r}_{2})\beta^{(2)}\Gamma_{\mu}^{(2)}
\label{resint2}%
\end{equation}
where the propagator $G(\boldsymbol{r}_{1},\boldsymbol{r}_{2})$ has Yukawa
form for mesons with linear couplings, and has to be evaluated numerically in
the other cases. We first concentrate on mesons with linear couplings. In this
case the propagator $G_{m}(\boldsymbol{r}_{1}-\boldsymbol{r}_{2})$ depends
only on $\boldsymbol{r}_{1}-\boldsymbol{r}_{2}$ and can be written in Fourier
space as
\begin{equation}
G_{m}(\boldsymbol{r}_{1}-\boldsymbol{r}_{2})=%
{\displaystyle\int}
\frac{d^{3}q}{(2\pi)^{3}}e^{i\boldsymbol{qr}_{1}}\Delta_{m}(\boldsymbol{q}%
)e^{-i\boldsymbol{qr}_{2}}%
\end{equation}
with the meson propagator
\begin{equation}
\Delta_{m}(\boldsymbol{q})=\frac{1}{\boldsymbol{q}^{2}+m_{m}^{2}}
\label{propagmomem}%
\end{equation}
The interaction \eqref{resint2} has the form
\begin{equation}
\hat{V}_{m}^{{}}(1,2)=\mp%
{\displaystyle\int}
\frac{d^{3}q}{(2\pi)^{3}}\hat{Q}^{\mu}(\boldsymbol{q},1)\Delta_{m}%
(\boldsymbol{q})\hat{Q}_{\mu}^{\dagger}(\boldsymbol{q},2)
\label{def:separable}%
\end{equation}
with
\begin{equation}
\hat{Q}^{\mu}(\boldsymbol{q},1)=g_{m}\beta^{(1)}\Gamma^{\mu(1)}%
e^{i\boldsymbol{qr}_{1}}%
\end{equation}
For each $\boldsymbol{q}$, $\hat{Q}^{\mu}(\boldsymbol{q})$ is a one-body
operator in $r$-space and in the 4-dimensional Dirac-space defined by the
combined index $1=(\boldsymbol{r}_{1},\alpha_{1})$. This definition of the
operator $\hat{Q}^{\mu}$ is flexible enough to allow also applications of the
meson exchange model with density dependent coupling constants $g_{m}%
(\boldsymbol{r}\mathbf{)}$ $=$ $g_{m}(\rho(\boldsymbol{r}))$. In the present
investigation, however, we do not follow this avenue. Considering the
$q$-integral as a sum over discrete values in $q$-space, the interaction
\eqref{def:separable} is a sum of separable terms. The corresponding two-body
matrix elements can thus be expressed by the one-body matrix elements of the
operators $\hat{Q}^{\mu}(\boldsymbol{q})$. Using this form we can evaluate the
two-body matrix elements
\begin{equation}
\langle kl^{\prime}|\hat{V}_{m}^{ph}|k^{\prime}l\rangle=\int\frac{d^{3}%
q}{(2\pi)^{3}}\langle k|\hat{Q}^{\mu}(\boldsymbol{q})|k^{\prime}\rangle
\Delta_{m}(\boldsymbol{q})\langle l|\hat{Q}_{\mu}(\boldsymbol{q})|l^{\prime
}\rangle^{\ast} \label{res:q-integ}%
\end{equation}
with the single particle matrix elements
\begin{equation}
\langle k|\hat{Q}^{\mu}(\boldsymbol{q})|k^{\prime}\rangle=\int d^{3}%
r~g_{m}\bar{\psi}_{k}(\boldsymbol{r})\Gamma^{\mu}(\boldsymbol{r}%
)e^{i\boldsymbol{q}\boldsymbol{r}}\psi_{k^{\prime}}(\boldsymbol{r}).
\label{spmatelm}%
\end{equation}
For the case with axial symmetry, the evaluation of these matrix elements is
best accomplished in cylindrical coordinates (\ref{cylindrical-r}). In this
case one finds that the integrals over the azimuth angles in coordinate and
momentum space can be evaluated analytically. This leads to the selection rule
$\Omega_{k}-\Omega_{k^{\prime}}=\Omega_{l}-\Omega_{l^{\prime}}$ (details are
given in Appendix \ref{app:A}).

For non-linear meson couplings the propagator $G_{m}(\boldsymbol{r}%
,\boldsymbol{r}^{\prime})$ depends on both coordinates and therefore we find
in Fourier space the matrix $\Delta_{m}(\boldsymbol{q},\boldsymbol{q}^{\prime
})$, which is calculated numerically by matrix inversion. This leads to a
four-fold integral in momentum space for the evaluation of the two-body matrix
elements (\ref{res:q-integ}) (for details see Appendix \ref{app:B}).

Summarizing this section, one finds for the elements of the RPA matrix
(\ref{rpa:eqn})%
\begin{align}
A_{mi,nj} &  =(\varepsilon_{m}-\varepsilon_{i})\delta_{mn}\delta
_{ij}+\nonumber\\
&  +\int\frac{d^{3}q}{(2\pi)^{3}}\langle m|\hat{Q}^{\mu}(\boldsymbol{q}%
)|i\rangle\Delta_{m}(\boldsymbol{q})\langle n|\hat{Q}_{\mu}(\boldsymbol{q}%
)|j\rangle^{\ast}\\
B_{mi,nj} &  =\int\frac{d^{3}q}{(2\pi)^{3}}\langle m|\hat{Q}^{\mu
}(\boldsymbol{q})|i\rangle\Delta_{m}(\boldsymbol{q})\langle\bar{j}|\hat
{Q}_{\mu}(\boldsymbol{q})|\bar{n}\rangle^{\ast}%
\end{align}
where $|\bar{n}\rangle=T|n\rangle$ is the time-reversed state to $|n\rangle$.
Using the symmetry properties of the operators $\hat{Q}^{\mu}(\boldsymbol{q})$
one obtains
\begin{equation}
\langle\bar{j}|\hat{Q}_{\mu}(\boldsymbol{q})|\bar{n}\rangle=(-)^{S}\langle
n|\hat{Q}_{\mu}^{{}}(\boldsymbol{q})|j\rangle
\end{equation}
where $S$ is the spin of the exchanged meson, i.e. $S=0$ for scalar mesons and
the time-like part of vector mesons and $S=1$ for the spatial part of the
vector mesons.

\subsection{Matrix elements in the intrinsic and in the laboratory frame}

So far we have solved the relativistic RPA equations in the intrinsic frame.
Neither the basis states, the $ph$-states based on a deformed ground state,
nor the eigenstates $|\nu,K^{\pi}\rangle$ of the RPA equations in
Eq.~\eqref{def:RPA-wavefunctions} are eigenfunctions of the angular momentum
operators $\mathbf{J}^{2}$ and $J_{z}$ in the laboratory frame. We therefore
call the states $|\nu,K^{\pi}\rangle$ wave functions in the intrinsic frame.
In fact, they have little in common with the wave functions in the laboratory
frame, which have to be eigenstates of the angular momentum operators
$\mathbf{J}^{2}$ and $J_{z}$. In order to calculate matrix elements which can
be compared with experimental data we therefore have to project onto good
angular momentum, i. e. on eigen spaces of these operators $\mathbf{J}^{2}$
and $J_{z}$ in the many-body Hilbert space.

Using the projection operators defined in Ref. \cite{RS.80} we obtain the wave
functions
\begin{align}
|\nu,K,IM\rangle &  =\hat{P}_{MK}^{I}|\nu,K^{\pi}\rangle\\
&  =\frac{2I+1}{8\pi^{2}}%
{\displaystyle\int}
d\Omega\mathcal{D}_{MK}^{I\ast}(\Omega)\hat{R}(\Omega)|\nu,K^{\pi}%
\rangle\nonumber
\end{align}
where $\mathcal{D}_{MK}^{I\ast}(\Omega)$ are the Wigner functions
\cite{Ed.57}, the irreducible representations of the group O(3) of rotations
in 3-dimensional space. They depend on the Euler angles $\Omega$, and $\hat
{R}(\Omega)$ is an operator which rotates the intrinsic wave function
$|\nu,K^{\pi}\rangle$ by the Euler angles $\Omega$. The evaluation of matrix
elements in the many-body Hilbert space using a projected wave functions is a
rather complicated task. In involves in particular the calculations of the
overlap integrals $\langle\nu,K^{\pi}|\hat{R}(\Omega)\mathcal{O}\hat{R}%
(\Omega^{\prime})|\nu^{\prime},K^{\prime\pi}\rangle$. It has been found, that
for well deformed intrinsic wave functions these overlap integrals are sharply
peaked at $\Omega=\Omega^{\prime}$. Replacing this sharply peaked functions by
Gaussians with a rather small width and in the extreme limit of strong
deformation by $\delta(\Omega-\Omega^{\prime})$ one obtains the so-called
\textit{needle approximation }\cite{RS.80}. The overlap functions are sharply
peaked, in particular for systems with many particles, and therefore the
needle approximation is not only valid in cases of strong deformations in the
geometrical sense, but in general for heavy systems with normal deformations,
and in the classical limit even for a spherical shape with a well defined
orientation. Moreover, it can be shown that the results obtained with this
approximate projection (the needle approximation), are equivalent to the
results of the particle plus rotor model \cite{Boh.52} where the orientation
$\Omega$ of the intrinsic frame is used as a dynamical variable (for details
see appendix \ref{app:E}).

The evaluation of the matrix elements in the laboratory system reduces to the
calculation of products of specific intrinsic matrix elements and geometrical
factors. This leads to the following expression for the reduced matrix
elements
\begin{align}
\langle I_{f}K_{f}||  &  \hat{\mathcal{O}}_{\lambda}||I_{i}K_{i}%
\rangle=(2I_{i}+1)(2I_{f}+1)\label{rpa:needle}\\
&  \left[  \left(
\begin{array}
[c]{ccc}%
I_{i} & \lambda & I_{f}\\
K_{i} & \mu & K_{f}%
\end{array}
\right)  \langle K_{f}|\hat{\mathcal{O}}_{\lambda\mu}|K_{i}\rangle\right.
\nonumber\\
&  \left.  +(-1)^{I_{i}+K_{i}}\left(
\begin{array}
[c]{ccc}%
I_{i} & \lambda & I_{f}\\
\bar{K}_{i} & \mu & K_{f}%
\end{array}
\right)  \langle K_{f}|\hat{\mathcal{O}}_{\lambda\mu}|\bar{K}_{i}%
\rangle\right] \nonumber
\end{align}
where $\langle K_{f}|\hat{\mathcal{O}}_{\lambda\mu}|K_{i}\rangle$ is the
intrinsic matrix element of the multipole operator $\hat{\mathcal{O}}%
_{\lambda\mu}$ which is easily calculated with the help of Eq.~\eqref{rpa:obo}.

In the following we therefore have to distinguish matrix elements and
transition densities in the intrinsic frame calculated directly with the
solutions of the RPA equation and matrix elements and transition densities in
the laboratory system, which are obtained after angular momentum projection in
the needle approximation in Eq. (\ref{rpa:needle}).

\section{Strength functions and sum rules}

\label{sec:multipole}

Experimental nuclear spectra show in the continuum excitations as resonances
with finite width. Since the diagonalization of the RPA equations is done in a
discrete basis, we obtain discrete eigenstates $|\nu\rangle$. Using Eqs.
(\ref{rpa:obo}) and (\ref{rpa:needle}) we can calculate for each of them the
reduced transition matrix elements for specific multipole operators, as for
instance the reduced transition probabilities ($BEI$- and $BMI$-values) for
electric and magnetic transitions
\begin{eqnarray}
B(EI,0\rightarrow I,K,\omega_{\nu})&=&\left\vert
\langle\nu,I\vert\vert\hat{Q}_{IK}\vert\vert 0\rangle\right\vert^2\\
B(MI,0\rightarrow I,K,\omega_{\nu})&=&\left\vert
\langle\nu,I\vert\vert \hat{M}_{IK} \vert\vert 0\rangle\right\vert^2%
\end{eqnarray}
It is well known that the width cannot be described well within the
RPA approach discussed here. On one side we work in a discrete basis,
and therefore the continuum is not treated properly and the escape
width is not taken into account, and on the other side RPA itself is
a linear approximation. It does not contain the coupling to $2p2h$-
and more complicated configurations and therefore does not allow a
proper treatment of the decay width. Higher order correlations, for
example the coupling to low-lying collective phonons
\cite{KST.04,LR.06,LRT.07}, have to be included for this purpose. It
is, however, also known from spherical RPA calculations, that this
method is able to describe rather well the position of the resonances
and the strength of the transitions for given multipole operators,
i.e. the percentage of the sum-rule exhausted by a specific
resonance. To overcome the problem of the width, we adopt a
phenomenological concept and average the discrete RPA strength
distribution obtained from the solution of the RPA equations in a
discrete basis with a Lorentzian function of a given width $\Gamma$.
For the electric response we have:
\begin{equation}
R(E)=\sum_{\nu}B(EI,0\rightarrow I,K,\omega_{\nu})\frac{1}{\pi}\frac{\Gamma
/2}{(E-\omega_{\nu})^{2}+(\Gamma/2)^{2}} \label{def:lorentzian}%
\end{equation}
and a corresponding expression holds for the magnetic response. This results
in continuous strength functions which can be compared with experimental
spectra and sum rules. The knowledge of the sum rules is of special interest,
since they represent a useful test of the models describing the collective
excitations \cite{RS.80}. For example, the energy weighted sum rule (EWSR)
for a transition operator $\hat{\mathcal{O}}$ can be represented as a double
commutator
\begin{equation}
S_{1}=\langle 0| [ \hat{\mathcal{O}},[H,\hat{\mathcal{O}}] ] |
0\rangle
\label{strcom}%
\end{equation}
If one assumes a non-relativistic Hamiltonian, a local operator $\hat
{\mathcal{O}}$ and a local two-body interaction, only contributions from the
kinetic energy contribute and one can evaluate this sumrule in a model
independent way:
\begin{equation}
S_{1}=\frac{\hbar}{2m}\frac{(2\lambda+1)^{2}}{4\pi}Z\langle r^{2\lambda
-2}\rangle
\end{equation}
These classical values for the sum rules are only approximate estimates. In
practical calculations they may be enlarged by an enhancement factor due to
the velocity dependence and due to exchange terms of the nucleon-nucleon
interaction. It can be shown, that many of these sum rules apply also in RPA
approximation. In this manuscript We evaluate the EWSR in the interval
below 30~MeV excitation energy as
\begin{equation}
S_{1}=\sum_{\nu}\omega_{\nu}B(EI,0\rightarrow I,K,\omega_{\nu}) \label{ewsreq}%
\end{equation}
In particular, the Lorentzian function in Eq.~\eqref{def:lorentzian} is
normalized in such a way as to give the same EWSR as
calculated with the discrete response
\begin{equation}
S_{1}=\sum_{\nu}\omega_{\nu}B(EI,0\rightarrow I,K,\omega_{\nu})=\int E~R(E)dE
\label{ewsreq2}%
\end{equation}
Sum rules also offer the possibility of a consistent definition of the
excitation energies of giant resonances, via the energy moments of the
discrete transition strength distribution
\begin{equation}
m_{k}=\sum_{\nu}E_{\nu}^{k}B((E/M)I,K,\omega_{\nu}) \label{strmoment}%
\end{equation}
In the case $k=1$ this equation defines the energy weighted sum rule of
Eq.~\eqref{ewsreq}. If the strength distribution of a particular excitation
mode has a well pronounced and symmetric resonance shape, its energy is well
described by the centroid energy
\begin{equation}
\bar{E}=\frac{m_{1}}{m_{0}}%
\end{equation}
Alternatively, mean energies are defined as
\begin{equation}
\bar{E}_{k}=\sqrt{\frac{m_{k}}{m_{k-2}}}%
\end{equation}
where the difference between the values $\bar{E}_{1}$ and $\bar{E}_{3}$ can be
used as an indication of how much the strength distribution corresponding to
an excitation mode is actually fragmented. If the multipole response is
characterized by a single dominant peak, the two moments are equal, i.e.
$\bar{E}_{1}=\bar{E}_{3}$. In the relativistic approach, due to the
\textit{no-sea} approximation, the sum in Eq.~\eqref{strmoment} runs not only
over the positive excitation energies, but also includes transitions to the
empty states in the Dirac sea which contribute with negative terms to the sum.
As it was pointed out in \cite{PW.85,SRM.89,MFR.89}, for the EWSR the double
commutator of Eq.~\eqref{strcom} should vanish, and it is another good check
for the numerical implementation.

\section{Transition densities}

\label{sec:transition}

In order to have a intuitive picture of the nuclear excitations we investigate
in the following the time evolution of the baryon density. Let us consider the
baryon four-current operator in coordinate space
\begin{equation}
\hat{\jmath}^{\mu}(\boldsymbol{r})=\sum_{i}\gamma^{\mu}\delta(\boldsymbol{r}%
-\boldsymbol{r}_{i})
\end{equation}
with single-particle matrix elements in the Dirac basis
\begin{equation}
j_{kk^{\prime}}^{\mu}(\boldsymbol{r})=\bar{\psi}_{k}(\boldsymbol{r}%
)\gamma^{\mu}\psi_{k^{\prime}}(\boldsymbol{r}),
\end{equation}
which can be written as
\begin{equation}
\hat{\jmath}^{\mu}(\boldsymbol{r})=\sum_{kk^{\prime}}j_{kk^{\prime}}^{\mu
}a_{k}^{\dagger}a_{k^{\prime}}. \label{def:currentop}%
\end{equation}
In order to calculate its time evolution within the RPA approximation, and for
a particular excitation mode $\nu$, we use Eq.~\eqref{rpa:obo} and find
\begin{equation}
\delta j^{\mu}(\boldsymbol{r})=\sum_{mi}(j_{im}^{\mu}(\boldsymbol{r}%
)X_{mi}^{(\nu)}+{j_{mi}^{\mu}}(\boldsymbol{r})^{\ast}Y_{mi}^{(\nu)})
\label{def:transdenscurrent}%
\end{equation}
Thus, the total time dependent baryon four-current for a given excitation mode
$\nu$ with energy $\omega_{\nu}$ is
\begin{equation}
j^{\mu}(\boldsymbol{r},t)=j_{0}^{\mu}(\boldsymbol{r})+\delta j^{\mu
}(\boldsymbol{r})e^{-i\omega_{\nu}t}+\delta{j^{\mu}(\boldsymbol{r})}^{\ast
}e^{i\omega_{\nu}t}%
\end{equation}
In particular, the baryon density $\rho(\boldsymbol{r},t)=j^{0}(\boldsymbol{r}%
,t)$ can be written as
\begin{equation}
\rho(\boldsymbol{r},t)=\rho_{0}(\boldsymbol{r})+\delta\rho(\boldsymbol{r}%
)e^{-i\omega_{\nu}t}+{\delta\rho(\boldsymbol{r})}^{\ast}e^{i\omega_{\nu}t}
\label{def:tddensity}%
\end{equation}

Throughout the rest of this paper, all types of intrinsic transition densities
refer to the \textit{baryon} intrinsic transition density in coordinate space,
$\delta\rho(\boldsymbol{r})$, as defined by the zero's component of
Eq.~\eqref{def:transdenscurrent}. As discussed in the previous section, we
also have to distinguish the transition densities in the intrinsic system from
those in the laboratory frame.

In a classical system the transition density would describe the actual
movement of particles. In the quantum mechanical description one considers a
time-dependent wave packet and decomposes it into the contributions of the
different excited eigenstates of the system. The transition density is an
off-diagonal matrix element between the stationary ground state $|0\rangle$
and the excited eigen state $|\nu\rangle$ and it is regarded as a measure of
the contribution of the eigen state $|\nu\rangle$ of the system to the
evolution of the time-dependent wave packet. To what extent an state
$|\nu\rangle$ can be interpreted in the classical sense depends on percentage
of the sum rule exhausted by the transition strength of this excitation mode
$|\nu\rangle$. Therefore the transition densities provide an intuitive
understanding of the nature of the excitation modes, It can be used as well
for the calculation of transition probabilities as for a qualitative
understanding of these modes.

In an axial symmetric system the transition density
in~\eqref{def:transdenscurrent} can be written as
\begin{equation}
\delta\rho(\boldsymbol{r})=\delta\rho(r_{\perp},z)e^{-iK\varphi}%
\end{equation}
where $K$ is the angular momentum projection of the excitation mode under
study. Note, that we distinguish in this section the coordinates $r_{\perp}$
(the distance from the symmetry axis) and $r$ (the distance from the origin).
Substituting this last expression in Eq.~\eqref{def:tddensity} we arrive at
\begin{equation}
\rho(r_{\perp},\varphi,z,t)=\rho_{0}(r_{\perp},z)+\left[  \delta\rho(r_{\perp
},z)e^{-i(K\varphi+\omega_{\nu}t)}+h.c\right]  \label{def:tddensityaxi}%
\end{equation}
The two dimensional quantities $\delta\rho(r_{\perp},z)$ will be plotted when
discussing intrinsic transition densities, and no further reference to the
phase expressed in the exponentials will be made. To interpret these plots, it
is useful to keep in mind that $\delta\rho(r_{\perp},z)$ has to be considered
together with Eq.~\eqref{def:tddensityaxi} in order to obtain the full three
dimensional geometrical picture. However, to be able to compare with
experimental transition densities measured in the laboratory frame, we project
the two dimensional intrinsic transition densities $\delta\rho(r_{\perp},z)$
on to good angular momentum. This is done by expanding the current operator in
Eq.~\eqref{def:currentop} in spherical coordinates $\boldsymbol{r}%
=(r\sin\theta\cos\varphi,r\sin\theta\sin\varphi,r\cos\theta)$ using the set of
spherical harmonics $Y_{LM}(\theta\varphi)$ as a basis
\begin{equation}
\hat{\jmath}^{\mu}(\boldsymbol{r})=\sum_{LK,kk^{\prime}}j_{L,kk^{\prime}}%
^{\mu}(r)Y_{LK}(\theta\varphi)\;a_{k}^{\dagger}a_{k^{\prime}}%
\end{equation}
where
\begin{equation}
j_{L,kk^{\prime}}^{\mu}(r)=\int d\cos\theta d\varphi\;j_{kk^{\prime}}^{\mu
}(\boldsymbol{r})Y_{LK}^{\ast}(\theta\varphi).
\end{equation}
The projected transition density reads
\begin{equation}
\delta\rho(\boldsymbol{r})=\delta\rho_{L}(r)Y_{LK}(\theta\varphi)
\end{equation}
with the radial projected transition density
\begin{equation}
\delta\rho_{L}(r)=\int d\cos\theta d\varphi\;\delta\rho(r_{\perp}%
,z)Y_{LK}^{\ast}(\theta\varphi) \label{def:prjtrdeq}%
\end{equation}
In the following figures we present the quantity $r^{2}\delta\rho_{L}$.
Because of the approximations in the derivation of Eq.~\eqref{rpa:needle},
this last equation only holds approximately. Nevertheless, we will see that
the results for well deformed nuclei are excellent. For example, the
transition density patterns for the Giant Dipole Resonances (GDR) and the
Pygmy Dipole Resonances (PDR), are in reasonable agreement with those found
experimentally and in other theoretical RPA studies in spherical symmetry.

\section{Spurious modes and numerical implementation}

\label{sec:spurious}

If the generator for a symmetry operation of the full two-body Hamiltonian,
which is represented by a one-body operator, does not commute with the ground
state density, there exists a Goldstone mode, a so-called spurious solution,
of the RPA equations with zero excitation energy associated with this
symmetry. These solutions are not really spurious, but they correspond to a
collective motion without a restoring force \cite{TV.62} and therefore they do
not correspond to oscillations with small amplitude. Examples are translations
or rotations. In principle this should not be significant, since we are
concerned only with the intrinsic structure of the nucleus. Thouless found
that for the exact solution of the self-consistent RPA equations these
spurious modes are orthogonal to all the other modes. They do not mix with
them and can be separated \cite{Tho.61}.

In practical applications, however, in many cases the spurious solutions are
not completely orthogonal to the physical states by various reasons. One
should be able to distinguish them from the true vibrational response of the
nucleus, as experience shows that this mixture can lead to seriously
overestimations in the strength distributions.

\begin{figure}[ptbh]
\includegraphics[width=8.6cm]{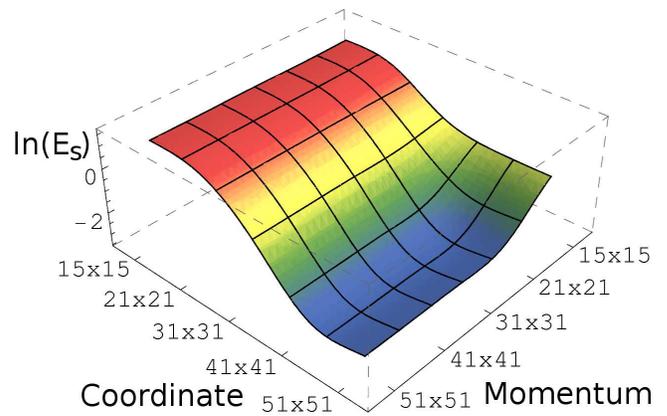}\caption{(Color online) Dependence of
the $K^{\pi}=1^{+}$ rotational spurious mode on the coordinate and momentum
mesh size for the non-linear model NL3 parametrization. For a coordinate and
momentum mesh size of (41x41) and (31x31), respectively, the accuracy limit of
the diagonalization procedure is achieved. The logarithmic scale in the
$z$-axis is used to enhance the readability of the graph. The lowest $z$ value
corresponds to a value of around 0.05~MeV.}%
\label{fig:spNe20M1mesh}%
\end{figure}

In normal calculations, because of numerical inaccuracies, truncation of the
$ph$-space and inconsistencies among the ground state and RPA equations, the
spurious states are often located at energies somewhat higher than zero and
often cause a mixing with physical states. There are several approaches to
overcome this problem. Some authors adjust a free parameter of the residual
interaction until the energy of the spurious mode goes to zero. Another method
is to remove \textit{a posteriori} the spurious components from the physical
states by projection. This is possible, because the wave functions of the
spurious modes are given by the matrix elements of the corresponding
generators \cite{RS.80}.

In this investigation a fully self-consistent implementation of the RPA is
used, and thus as long as numerical inaccuracies are kept to a minimum, the
spurious modes decouple without further complications. Because the block-wise
structure of the RPA matrix, they are expected to be present only for specific
quantum numbers when specific symmetry constraints are met; since we are
restricting to axial symmetry, their expected appearance can be summarized as

\begin{itemize}
\item A rotational spurious mode for the $K^{\pi}=1^{+}$ channel associated
with rotations of the nucleus as a whole around an axis perpendicular of the
symmetry axis in $z$-direction. Its generator is the angular momentum operator
$\hat{J}_{+}=\hat{J}_{x}+i\hat{J}_{y}$ \cite{Mar.77b}.

\item A translational spurious mode for the $K^{\pi}=0^{-},1^{-}$ channels
associated with the translation of the nucleus as a whole. Its generators are
the linear momentum operators $\hat{P}_{z\text{ }}$and $\hat{P}_{+\text{ }}$.
\end{itemize}

\begin{figure}[ptbh]
\includegraphics[width=8.6cm]{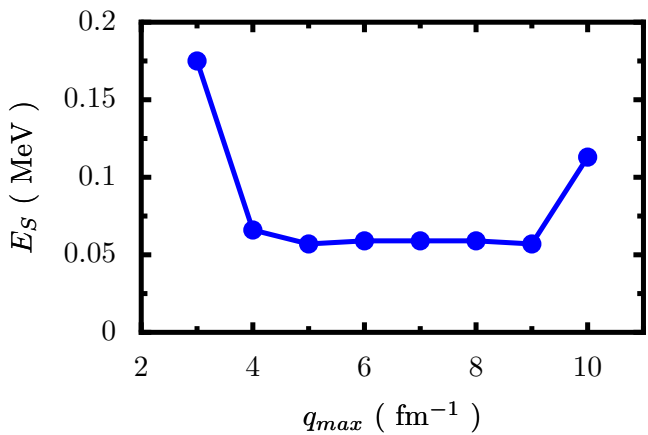}\caption{(Color online) Spurious
$K^{\pi}=1^{+}$ rotational mode dependence on the maximum interaction momentum
while keeping the number of mesh points constant. Good numerical results for a
momentum mesh size of (31x31) can be achieved with a maximum momentum in the
interval $5 < q_{max} < 9$.}%
\label{fig:spNe20M1momentum}%
\end{figure}

In fact, the position of the spurious modes provides a very accurate test of
the actual numerical implementation of the RMF+RRPA framework. Thus, it is
important to study their evolution with the approximations performed. In the
present status of the implementation seven parameters control the numerical
accuracy, and can be categorized in two groups. The first group specifies the
precision of the numerical integrations. In this category are included the
number of coordinate and momentum lattice points and the upper boundary of the
momentum integrals. The second group deals with the size of the configuration
space, and includes the energy cutoffs for $ph$- and $ah$-pairs.

Since it is unfeasible to study this seven-dimensional surface in detail, when
studying the dependence of the spurious modes on one, or a set of, parameters,
those not under scrutiny were fixed to best possible values the hardware
supports. This means, in particular, that the full $ph$ configuration space is
taken if not otherwise stated, and that the maximum momentum is fixed to
$q_{\max}=8$~fm$^{-1}$, well above the Fermi momentum of the nucleus.

In Fig.~\ref{fig:spNe20M1mesh} the position of the rotational $K^{\pi}=1^{+}$
spurious mode in $^{20}$Ne is plotted against the number of points in the
coordinate and momentum lattices. For a relatively low number of points a
plateau is reached where further improvement of the accuracy cannot be
achieved. The optimal number of evaluation points for the quadratures is
therefore around 41x41, which allows for very precise calculations.
Furthermore, additional tests show that the overall precision in the
determination of the energy of excited states of the code is capped out at
0.05~MeV, which is surprisingly good.

\begin{figure}[ptbh]
\includegraphics[width=8.6cm]{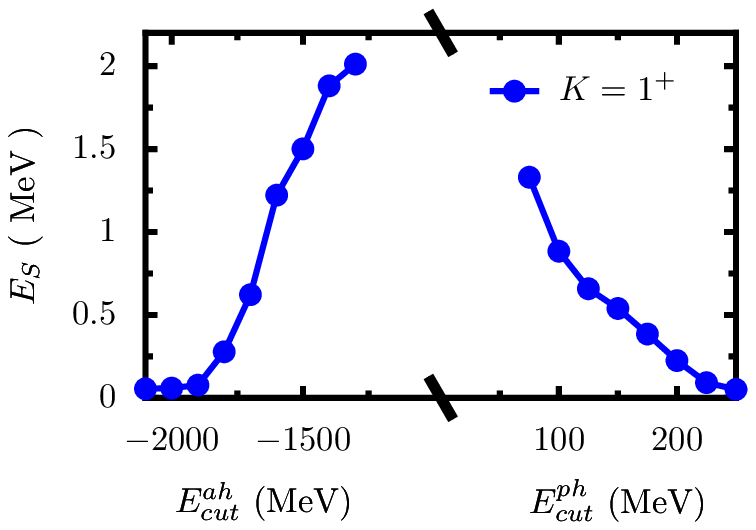} \newline%
\includegraphics[width=8.6cm]{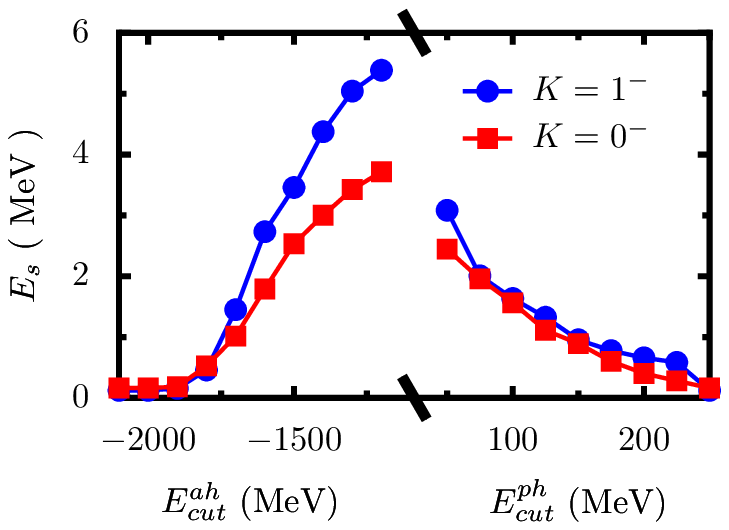} \caption{(Color online) On the upper
plot we show the dependence of the $K^{\pi}=1^{+}$ rotational spurious mode on
the size of the $ph$- and $ah$-space size. The lower plot shows the same
dependence for the $K^{\pi}=1^{+}$ and $K^{\pi}=0^{-}$ translational spurious
modes.}%
\label{fig:spNe20ahph}%
\end{figure}

Figure~\ref{fig:spNe20M1momentum} depicts the position of the rotational
$K^{\pi}=1^{+}$ spurious mode for $^{20}$Ne against the maximum momentum of
the expansion used in the integral for the evaluation of the single particle
matrix elements in Eq.~\eqref{spmatelm}. The flat region between 5 and
9~fm$^{-1}$ hints that a maximum momentum of $q_{\max}=5$ fm$^{-1}$ provides
enough precision for the proper decoupling of the spurious mode. The increase
observed in the position of the spurious mode for maximum momentum values
larger than 9~fm$^{-1}$ is an artifact due to the number of points for the
momentum lattice being fixed at (31x31).

\begin{figure}[ptbh]
\includegraphics[width=8.6cm]{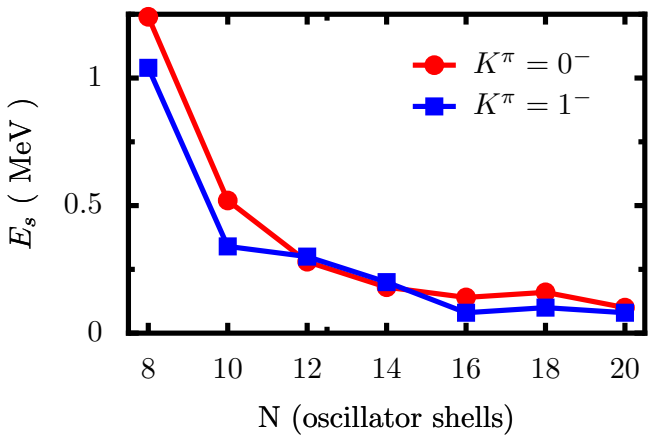}\caption{(Color online) Dependence of
the $K^{\pi}=1^{-}$ and $K^{\pi}=0^{-}$ translational spurious modes on the
configuration space size, dictated by the number of oscillator shells used in
the ground-state calculation.}%
\label{fig:spNe20E1shells}%
\end{figure}

Figure~\ref{fig:spNe20ahph} shows the dependence of the rotational $K^{\pi
}=1^{+}$ and translational $K^{\pi}=0^{-},1^{-}$ spurious modes on the
configuration space size for $^{20}$Ne, as calculated with the NL3 parameter
set. In the translational case two curves are plotted, one for the $K^{\pi
}=0^{-}$ mode and one for the $K^{\pi}=1^{-}$ mode. It is interesting to note
that, even if the spurious mode can be brought very close to zero, it requires
the inclusion of almost all the possible $ph$-pairs in the configuration
space. In this specific case, i.e. in $^{20}$Ne, that amounts to the inclusion
of roughly five thousand pairs. Several test indicate that the situation
improves greatly in heavier nuclei, where usually 5\% percent of all possible
$ph$-pairs are enough to decouple the spurious modes at energies around 0.5~MeV.

\begin{figure}[ptbh]
\includegraphics[width=8.6cm]{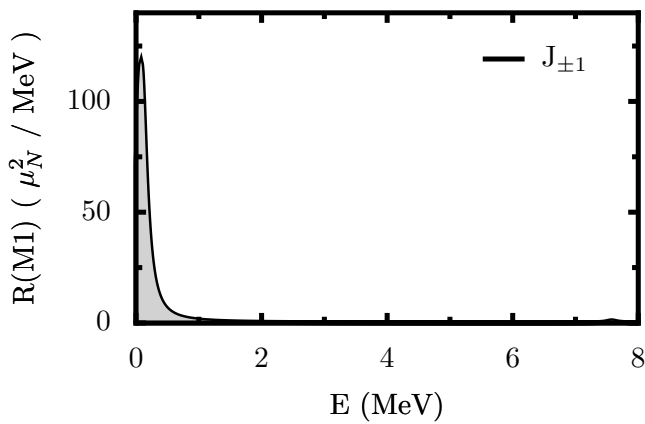} \caption{(Color online) Response in
$^{20}$Ne to the operator $J_{\pm1}$, generator of rotations around a
perpendicular of the symmetry axis. Almost 100\% of the strength is exhausted
by the spurious mode (situated at 0.08~MeV), with minimal admixture to the
physical states.}%
\label{fig:Ne20-Jx}%
\end{figure}

Since for the solution of the ground-state the equations of motion are
expanded in a harmonic oscillator basis, the configuration space where the RPA
is solved does not spawn the whole Hilbert space, even if all possible
$ph$-pairs are taken. The quality of this expansion depends on the number of
major oscillator shells used, and the results obtained at the RPA level will
be influenced by this approximation. In particular, the proper decoupling of
the translational spurious mode is very sensitive to the number of oscillator
shells employed. In Fig.~\ref{fig:spNe20E1shells} the translational spurious
mode is plotted versus the number of oscillator shells used in the ground
state. Already with the inclusion of 16 major shells is enough to achieve a
precision in the spurious mode of around 0.1~MeV. In all practical cases
presented in the this study the number of oscillator shells was chosen between
12 and 16, depending on the desired final precision and the availability of
computer resources.

However, more important that the actual position of the spurious modes is
their admixture to the real physical states. For the same reasons the spurious
mode does not appear at exactly zero energy, the physical states are not
completely orthogonal to it, producing unreal results and very often
overestimated strength.

Moreover, there is one important property of the spurious modes that can also
be used to measure the extent of their admixture with the rest of the RPA
states. They are not normalizable in the sense of Eq.~\eqref{rpa:rpanorm}
because
\begin{equation}
X_{mi}=Y_{mi} \label{spmode:xy}%
\end{equation}
However, in all numerical implementations the relation \eqref{spmode:xy} is
only approximately fulfilled because the spurious modes do not decouple
exactly. How good the decoupling of the spurious modes is can be measured
comparing the relative norms of the different eigen-modes. For an approximate
spurious mode labeled as $(sp)$ it should hold that $X_{mi}^{(sp)}\approx
Y_{mi}^{(sp)}$, or
\begin{equation}
\delta:=1-\frac{\sum_{mi}|Y_{mi}^{(sp)}|^{2}}{\sum_{mi}|X_{mi}^{(sp)}|^{2}}%
\ll1 \label{relnormsp}%
\end{equation}
while for any other RPA mode $\nu$, by initial assumption, it holds that
$X_{mi}\gg Y_{mi}$, and thus
\begin{equation}
\delta:=1-\frac{\sum_{mi}|Y_{mi}^{(sp)}|^{2}}{\sum_{mi}|X_{mi}^{(sp)}|^{2}%
}\approx1
\end{equation}
Our tests indicate that when the spurious modes are located below 0.5~MeV, the
value of $\delta$ in Eq.~\eqref{relnormsp} is at least three orders of
magnitude smaller than the ones belonging to normal RPA modes. This is a very
good indicative of the proper decoupling of the Goldstone modes.

\emph{ }As an example of the low admixture of spurious components with the
physical states, Figure~\ref{fig:Ne20-Jx} shows the response to the generator
of the rotational spurious mode, the operator $J_{\pm1}$, which represents
rotations around an axis perpendicular to the symmetry axis. More than 99.99\%
of the strength is exhausted by the spurious mode, which is located below
0.1~MeV. Similar results are obtained for the translational spurious modes in
$^{20}$Ne.

In general, it was observed that, if the position of the spurious mode is
below 1~MeV, the strength function of the rest of the spectrum is mostly
unaffected. The spectrum in the low energy region, below 5~MeV, is, however,
more sensitive to admixtures of the spurious modes; as a rule of thumb, the
confidence limit in the position of the spurious mode, for a proper
decoupling, has been consistently found around 0.5~MeV.

\begin{figure}[ptbh]
\includegraphics[width=8.6cm]{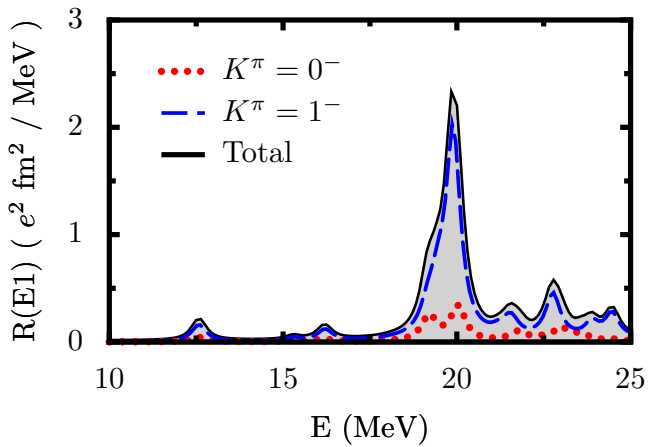}\caption{(Color online) $K^{\pi}=0^{-}$
and $K^{\pi}=1^{-}$ response to the E1 transition operator for the spherical
nucleus $^{16}$O.}%
\label{fig:O16}%
\end{figure}

There is still another test that can be devised in order to check the
consistency of the whole framework, namely the conservation of spherical
symmetry. Even though all the formulas are particularized to the case of axial
symmetry, the interaction is rotationally invariant, so they should still be
valid when a spherical ground-state is taken as basis for the RPA
configuration space, i.e., they should preserve spherical symmetry.

In Figure~\ref{fig:O16} is plotted the E1 excitation strength for the
spherical nucleus $^{16}$O. Since the E1 operator is a rank-one tensor, it has
three possible angular momentum projections, $K=-1,0,1$, that have to be
calculated separately. The response for the modes with $K=-1$ and $K=1$ are
identical and correspond to vibrations perpendicular to the symmetry axis,
i.e. one can calculate only one of them and double its contribution. The $K=0$
mode corresponds to vibrations along the symmetry axis. If the nucleus is
prolate, like $^{20}$Ne, the response for in the $K=0$ mode should lie at
lower energies than the $K=1$ mode, as the potential is flatter in the
direction of the symmetry axis. However, if the nucleus is spherical, like
$^{16}$O, there is no distinction between the $K=0$ and $K=1$ modes, and their
corresponding excitation strength should lie at exactly the same energies.
From Figure~\ref{fig:O16} one can attest that the procedure for the solution
of the RPA equation in axial symmetry indeed preserves rotational symmetry
with a good degree of accuracy.

The study presented concerning the decoupling of the spurious modes and the
preservation of spherical symmetry shows that the numerical implementation
solves the equations posed by the self-consistent RMF+RPA framework in axial
symmetry. We have also ascertained that a high degree of accuracy can be
achieved in real calculations, as well as validated the good reproduction of
formal and mathematical aspects of the RPA theory.

\section{Applications in $^{20}$Ne}

\label{sec:applications}

As a first application of the RMF+RRPA framework we have undertaken a model
study of the magnetic and the electric dipole response in $^{20}$Ne. This
nucleus offers several advantages. Its ground state is well deformed and
exhibits a prolate shape in the RMF model, with a quadrupole deformation
parameter $\beta\approx0.5$. Another advantage is the reduced number of
nucleons to be taken into account in the calculations, which translates in
fast running times and thus in the possibility of detailed analysis. With the
optimal number of oscillator shells for a ground state calculation with full
precision, the number of pairs never exceeds five thousand. Furthermore,
because the number of protons and neutrons is the same, switching off the
electromagnetic interaction should give identical results for both protons and
neutrons. Using this technique, very detailed checks can be carried out on the
isospin part of the interaction, and its consistency can be further
established. All these reasons make $^{20}$Ne the ideal theoretical playground
where to introduce the concepts that can later be used in the study of more
complex systems. In the this section we shall present two sample applications
for the well deformed nucleus $^{20}$Ne.

\subsection{The magnetic dipole (M1) response}

We first consider the magnetic dipole response. The discovery of low-lying M1
excitations, known as \textit{scissors mode}, was made by Richter and
collaborators in $^{156}$Gd in Darmstadt through a high-resolution inelastic
electron scattering experiment \cite{BRS.84}. The search for such a mode was
stimulated by the theoretical prediction of a collective mode, where the
deformed proton distribution oscillates in a rotational motion against the
deformed neutron distribution
\cite{SR.77,IP.78a,IP.78b,Hil.84,Hil.92}. The name \textquotedblleft scissors
mode\textquotedblright\ was indeed suggested by such a geometrical picture. An
excitation of similar nature was also predicted by group-theoretical models
\cite{Iac.81, Die.83,Iac.84}.

The mode has been detected in most of the deformed nuclei ranging from the
$fp$-shell to rare-earth and actinide regions. The mode has been well
characterized, and it is established that it is fragmented over several
closely packed M1 excitations. For reviews to this mode and for recent
semiclassical investigations see Refs.
\cite{VHJ.86,KPZ.96,Rich.95,Zaw.98,BalS.07a,BalS.07b}

A byproduct of the systematic study of the \textit{scissors mode} was the
discovery of spin excitations. Inelastic proton scattering experiments on
$^{154}$Sm and other deformed nuclei found a sizable and strongly fragmented
M1 spin strength distributed over an energy range of 4-12 MeV \cite{FWR.90}.
The experimental discovery stimulated theoretical investigations in the RPA
approximation \cite{HHR.98}.

\begin{figure}[ptbh]
\begin{center}
\includegraphics[width=8.6cm]{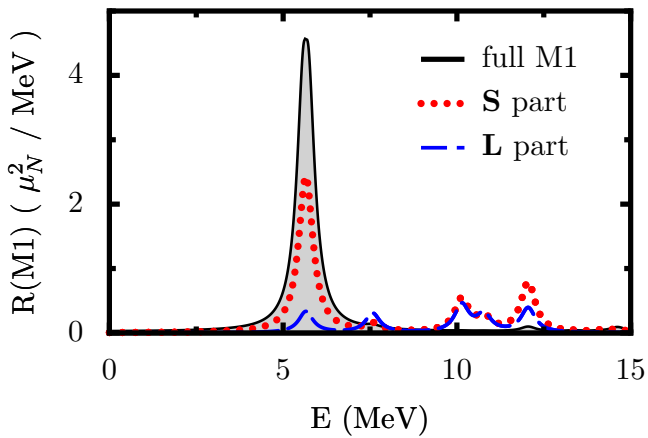} \newline
\end{center}
\caption{(Color online) M1 Excitation strength for $^{20}$Ne, using the NL3
parameter set. A very well developed peak can be seen around 5.8~MeV. Its
structure is composed mostly of spin flip transitions.}%
\label{fig:Ne20M1}%
\end{figure}

Figure \ref{fig:Ne20M1} shows the response to the M1 magnetic dipole operator
in the nucleus $^{20}$Ne . The shaded region corresponds to the full M1
response; the blue dashed line is the response to orbital part of the M1
operator and the red dotted line refers to the spin part. The calculations
were performed with the maximum precision allowed by the current
implementation of the computer code. The number of pairs is around five
thousand. Optimal numerical parameters were chosen in order to minimize the
error. The rotational spurious mode is well separated, situated around
0.1~MeV, i.e. no admixture with the vibrational response is observed.

Only one prominent peak is found around 5.7~MeV. Regrettably, there is no
experimental data available for the magnetic response in this nucleus.
Theoretical studies using large scale shell model calculations \cite{CP.86}
predict a low lying orbital mode around 11~MeV for $^{20}$Ne, in strong
disagreement with our results. However, other calculations \cite{LZ.87}
performed in $^{22}$Ne, with the same shell model interaction, exhibit two
dominant low lying peaks around 5-6~MeV. The orbital contribution to the total
response is below 25\%, which is in better agreement with results found within
our RMF+RRPA calculations, where fragmented strength with similar
characteristics is found in the same energy region.

Regarding the contributions from the orbital and spin components of the M1
operator to the total response, it can be observed in Figure \ref{fig:Ne20M1}
that there are two differentiated energy regions. Around the main excitation
peak at 6~MeV there is an enhancement of the response due to the additive
interference of the orbital and spin contributions. In contrast, in the energy
region above 6.5~MeV we observe the opposite, destructive interference and
both contributions cancel. This feature of the M1 strength distribution has
been also found in other studies \cite{HHR.98} and is much more evident in the
case of heavier nuclei.

From this figure we can also recognize that the main contribution to the total
response comes from spin excitations. The supposed orbital character of the
low lying spectra in the M1 transitions is eclipsed by the preponderance of
spin flip strength, three times larger than the orbital response. Again, this
is in disagreement with the cited shell model calculation \cite{CP.86}, which
in $^{20}$Ne predicts a much bigger orbital contribution to the total
strength. However, low lying collective transitions in such a light nucleus as
$^{20}$Ne cannot be expected to be exceptionally well described by the
RMF+RRPA theory. In few nucleon systems, the single particle structure around
the Fermi surface is of the utmost importance in the calculation of low-lying
excitations. As such, the results produced in a self-consistent mean field
calculation are not so reliable. A better description would require a proper
account of excitations to the continuum above the coulomb barrier and probably
for higher order correlations at the time dependent mean field level. The
situation improves in heavier nuclei, were mean field theories were designed
to yield good results at low computational costs.

\begin{table}[ptbh]
\begin{center}
\begin{tabular}
[c]{lrlcrc}%
\multicolumn{5}{l}{Peak at 5.7~MeV} & \\
&  &  &  &  & $\varepsilon_{1}-\varepsilon_{2}$\\\hline
P & 49\% & $\frac{1}{2}^{+}[220]$ & $-$ & $\frac{3}{2}^{+}[211] $ & 5.15\\
N & 48\% & $\frac{1}{2}^{+}[220]$ & $-$ & $\frac{3}{2}^{+}[211] $ & 5.22\\
P & 1\% & $\frac{1}{2}^{+}[220]$ & $-$ & $\frac{1}{2}^{+}[211] $ & 9.73\\
N & 0.9\% & $\frac{1}{2}^{+}[220]$ & $-$ & $\frac{1}{2}^{+}[211] $ &
10.17\\\hline
\end{tabular}
\end{center}
\caption{$ph$ structure for the 5.7~MeV M1 transition mode in $^{20}$Ne for
the NL3. The second column refers to the normalization of the RPA amplitudes.
The level quantum numbers in the third column are $\pm\Omega^{\pi}$, where
$\pm\Omega$ is the angular momentum projection over the symmetry axis and
$\pi$ is the parity. In square brackets are the quantum numbers of the
oscillator state which contributes most to the mean field single particle
level. The effect of coulomb interaction can be seen as the small differences
in the mixing percentages for protons and neutrons. A calculation with the
electromagnetic interaction switched off gives as a result a perfect isospin
symmetry, with no differences observable within the accuracy of the computed
results.}%
\label{m1:tbl:ne20}%
\end{table}

Nevertheless, it is still interesting to delve further into the study of the
properties of the main excitation peak, as the same analysis can be performed
in heavier nuclei and many of the general features will still be present. The
study of the structure of the excitation peaks can be carried out in detail
attending to their $ph$ structure. The contribution $C_{ph}$ from a particular
proton or neutron $ph$ configuration to a RPA state is determined by
\begin{equation}
C_{ph}=(|X_{ph}^{\nu}|^{2}-|Y_{ph}^{\nu}|^{2}) \label{phcontrib}%
\end{equation}
where $X^{\nu}$ and $Y^{\nu}$ are the RRPA amplitudes associated with a
particular excitation energy. Table \ref{m1:tbl:ne20} outlines the single
particle decomposition of the dominant M1 peak observed in
Figure~\ref{fig:Ne20M1}. All the strength is provided by a single particle
transition within the sd-shell, from the last level in the Fermi sea to the
first consecutive unoccupied level. The low collectivity indicates that,
within the RMF+RRPA model, the spectrum of the M1 operator in $^{20}$Ne is of
single particle character. Each of the two Dirac spinors in the particle-hole
pair corresponds to a eigen state of the static RMF potential. They can be
characterized by the Nilsson quantum numbers $\Omega^{\pi}[Nn_{z}\Lambda]$ of
their largest component in an expansion in anisotropic oscillator wave
functions. Here $\Omega$ is total angular momentum projection onto the
symmetry axis, $\pi$ is the parity, $N=2n_{r}+n_{z}+\Lambda$ is the major
oscillator quantum number, and $\Lambda=\Omega-m_{s}$ is the projection of the
orbital angular momentum on to the symmetry axis. From these quantum numbers
one concludes the following approximate selection rules: $\Delta\Omega=+1$,
$\Delta N=0$, $\Delta n_{z}=-1$ and $\Delta\Lambda=+1$. The orbital character
of the excitation peak is confirmed by the fact that $\Delta\Omega
=\Delta\Lambda$, which implies that the change in the magnetic quantum number
$m_{s}$ is zero. It is interesting to note that even if the approximate
selection rule points to an orbital character for the mode, the spin strength
is nevertheless dominant.

\begin{figure}[ptbh]
\begin{center}
\includegraphics[width=8.6cm]{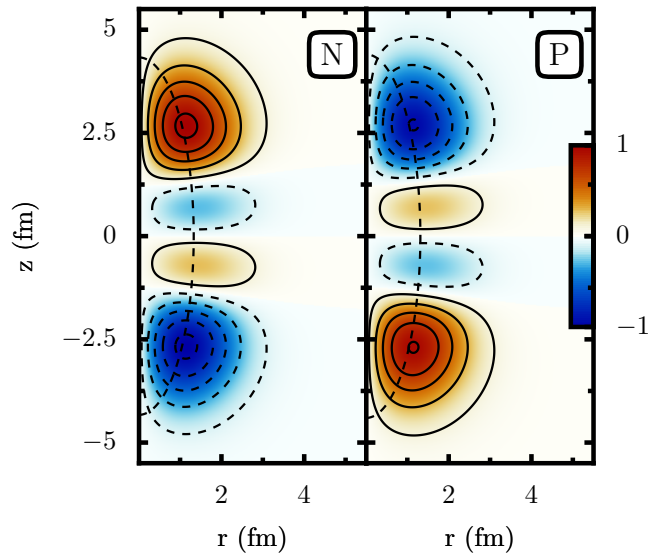}
\end{center}
\caption{(Color online) Intrinsic transition density for neutrons (left) and
protons (right) of the M1 peak at 5.7~MeV in in $^{20}$Ne. Full (red shade)
and dashed lines (blue shade) indicate positive and negative values,
respectively. The $z$-coordinate runs along the symmetry axis and $r$ is the
distance from the symmetry axis. The thin dotted line represents the rms
radius of the ground state neutron or proton density, and qualitatively marks
the position of the ground state nuclear surface.}%
\label{fig:Ne20M1rho}%
\end{figure}

\begin{figure}[ptbh]
\begin{center}
\includegraphics[width=8.6cm]{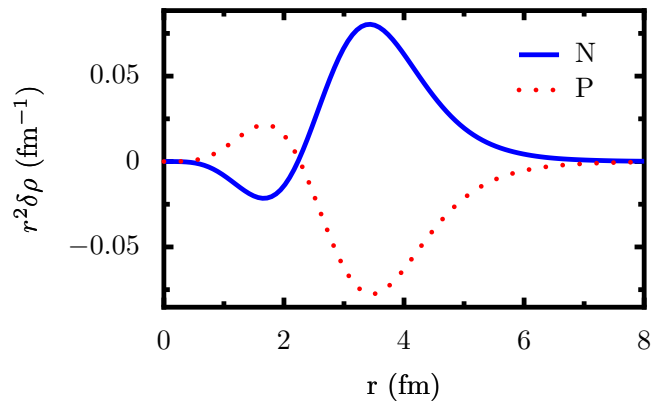}
\end{center}
\caption{(Color online) Radial part of the projected (to $I=1$, $M=1$)
transition densities of the M1 peak at 5.7~MeV. $r$ is the distance from the
symmetry axis. The prominent isovector nature is patent in the graphs.}%
\label{fig:Ne20M1prho}%
\end{figure}

It is difficult to form a geometrical image of the nature of an excitation,
having at hand only the information given in Table \ref{m1:tbl:ne20}. For that
purpose, it is always useful to compare the the neutron and proton intrinsic
transition densities in a plot. Figure \ref{fig:Ne20M1rho} shows a color plot
of the transition densities at an excitation energy of 5.7~MeV. Color is used
to indicate the value of the function, with blue for negative values and red
for positive ones. Regions with the same kind of line (solid or dashed), or
color shade (red or blue), for both protons and neutrons are indicative of an
in-phase vibration, while in regions where the opposite is true protons and
neutrons vibrate out-of-phase. In this case the excitation is of clear
isovector nature, and we can observe the typical structure of a scissors mode;
neutrons and protons are out of phase over the full space, with a
concentration near the caps of the prolate nuclear shape.

In such a simple case as the one found in $^{20}$Ne the interpretation of the
two dimensional color plot for the transition densities is very clear. They
represent the intrinsic transition densities, referred to the intrinsic frame
of reference, where only the total angular momentum projection on the symmetry
axis is well defined. In that regard, they are expected to contain admixtures
from all possible angular momenta. However, the transition operator (M1 in
this specific case) restricts the major contributions of the transition
densities to the total response to its own total angular momentum, i.e. to in
the case of M1 transitions to $I=1$. It is therefore advisable to project out
the weaker-contributing angular parts from the densities to obtain the actual
transition density that would be observed in the laboratory frame of
reference. For the M1 operator that means retaining , with the help of
Eq.~\eqref{def:prjtrdeq} only the contributions coming from angular momentum
$I=1$. In Figure \ref{fig:Ne20M1prho} the radial part of such a projected
transition density is plotted for the main peak in the $^{20}$Ne M1 response.

Both transition densities are almost the mirror of each other, a very clear
indication of the pure isovector nature of the mode at 5.7~MeV. We have
already seen that in simple geometrical terms this mode can be interpreted as
a rotation of neutrons against protons around an axis perpendicular of the
symmetry axis. Furthermore, details in Figure \ref{fig:Ne20M1rho} show that
two distinct regions can be distinguished. They are separated at around 2~fm
from the origin, where the direction of rotation for protons and neutrons
changes. The appearance of two regions (as depicted in Figure
\ref{fig:Ne20M1prho}) is already a strong hint that the simple picture of the
proton density rotating against the neutron density as rigid rotors (as in the
Two Rotor Model \cite{IP.78a}) does not reflect reality in this nucleus. The
traditional scissors picture considers the neutron and proton densities as the
blades of a scissor oscillating against each other. In addition, one has to
take into account the angular momentum inherent in a $K=1^{+}$ excitation: it
can be descibed as an oscillation of the scissor which rotates at the same
time rotating slowly around its longitudinal symmetry axis. However, the
picture we derive from the results of our calculation is somewhat different.

\subsection{The Electric dipole (E1) response}

\begin{figure}[pth]
\includegraphics[width=8.6cm]{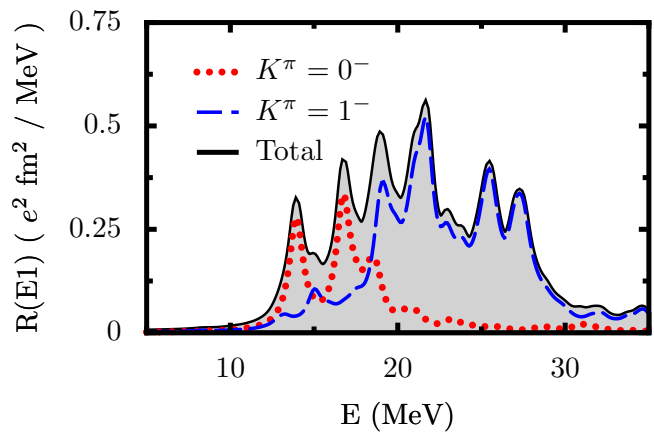}\caption{(Color online) E1 strength in
$^{20}$Ne, as calculated with the NL3 parameter set.}%
\label{fig:Ne20E1}%
\end{figure}

We have chosen the electric dipole response in $^{20}$Ne as a second example
application of the RMF+RRPA formalism with axial symmetry. In Fig.
\ref{fig:Ne20E1} the E1 strength is plotted as calculated with the NL3
parameter set. The red curve corresponds to excitations along the symmetry
axis with $K^{\pi}=0^{-}$, while the blue curve are those perpendicular to the
symmetry axis with $K^{\pi}=1^{-}$. In principle, for prolate nuclei, as is
the case for $^{20}$Ne, the strength due to the $K^{\pi}=0^{-}$ mode should
lie at lower energies compared to the $K^{\pi}=1^{-}$ mode. As the nuclear
potential must be flatter (more extended) along the symmetry axis, it is
energetically more favorable for the nucleons to oscillate in that direction
than along an axis perpendicular to the symmetry axis, where the nuclear
potential is narrower. It is possible, therefore, to relate the nuclear
deformation with the energy separation of the two modes \cite{Dan.58,Oka.58}.

\begin{figure}[ptbh]
\includegraphics[width=8.6cm]{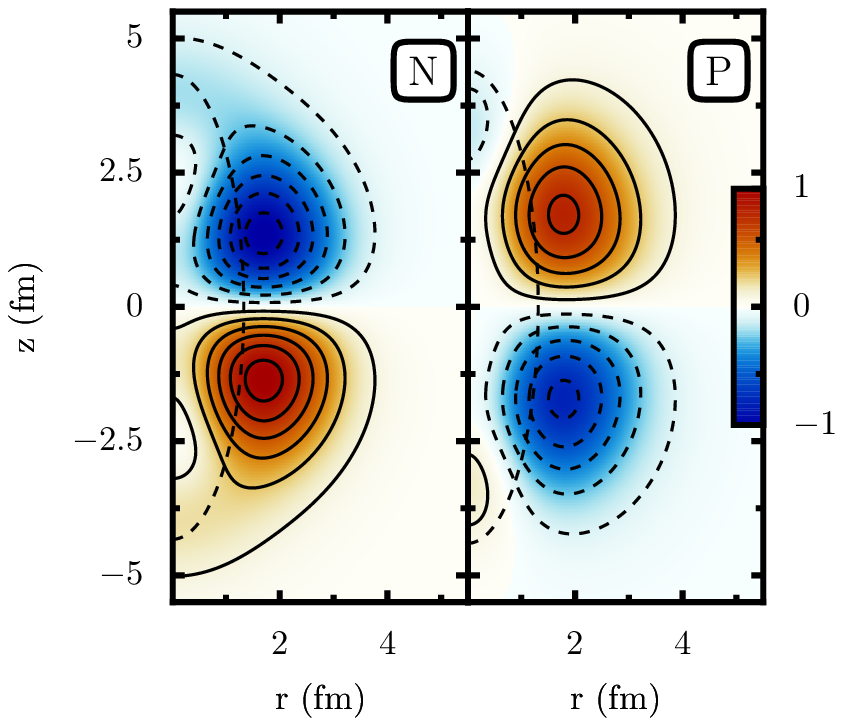} \newline%
\includegraphics[width=8.6cm]{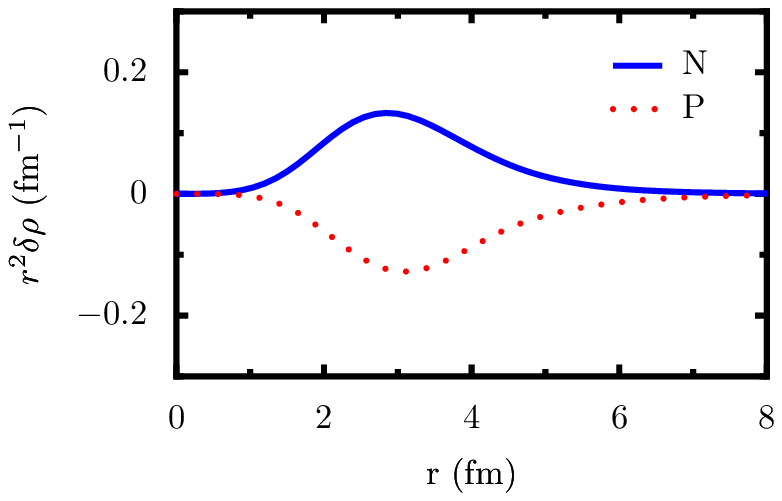}\caption{(Color online) $^{20}$Ne
IVGDR transition density for the $K^{\pi}=0^{-}$ peak at 16.7~MeV, NL3
parameter set.}%
\label{fig:Ne20E1rho1673}%
\end{figure}

\begin{figure}[ptbh]
\includegraphics[width=8.6cm]{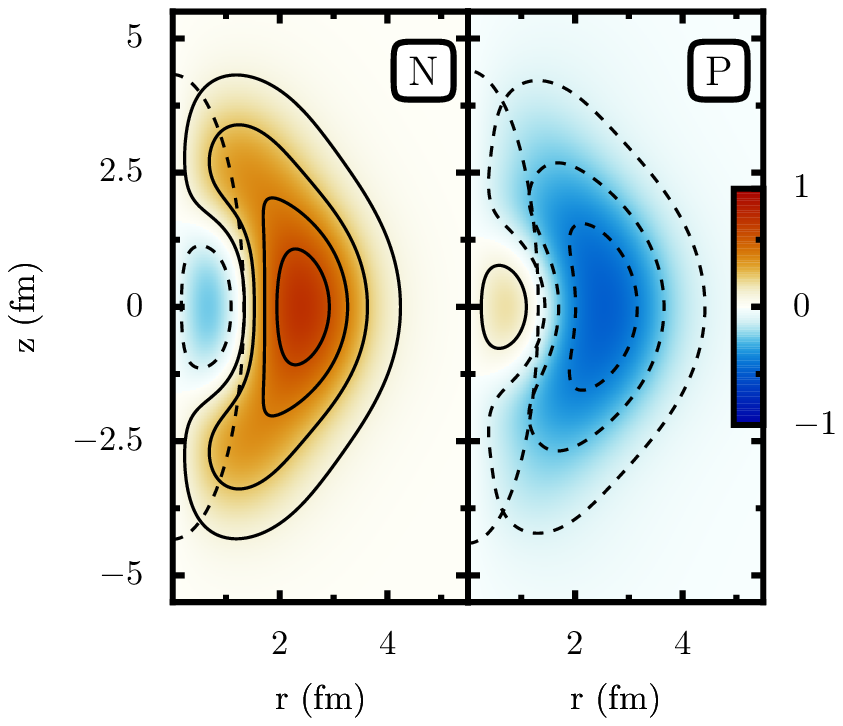} \newline%
\includegraphics[width=8.6cm]{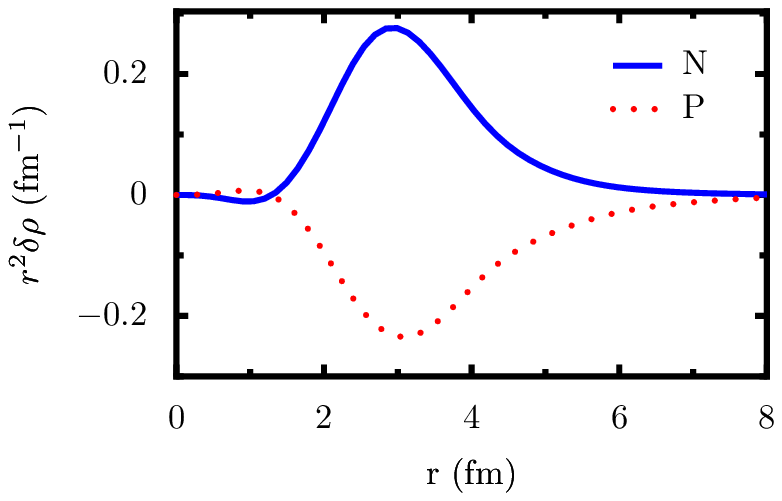}\caption{(Color online) $^{20}$Ne
IVGDR transition density for the $K^{\pi}=1^{-}$ peak at 21.3~MeV, NL3
parameter set.}%
\label{fig:Ne20E1rho2131}%
\end{figure}

The splitting of the response due to the broken spherical symmetry, and its
interpretation, can be observed in Figures~\ref{fig:Ne20E1rho1673} and
\ref{fig:Ne20E1rho2131}. The former is the transition density for the main
IVGDR peak at 16.73~MeV observed in the $K^{\pi}=0^{-}$ response, while the
latter corresponds to the peak at 21.31~MeV in the $K^{\pi}=1^{-}$ mode. The
prolate deformation is evident, as the intrinsic transition densities are
elongated in the direction of the $z$-axis. The character of the $K^{\pi
}=1^{-}$ mode as a vibration along a perpendicular of the symmetry axis is
easily recognizable in Figure~\ref{fig:Ne20E1rho2131}. By comparison, the
transition density in Figure~\ref{fig:Ne20E1rho1673} is then easily
interpreted as a vibration along the symmetry axis. As it is expected for the
IVGDR, the neutron-proton vibrations are out of phase over the same spatial
regions. It is more evident still looking at their respective projections to
the laboratory system of reference, which are shown in the lower plots of
Figures~\ref{fig:Ne20E1rho1673} and \ref{fig:Ne20E1rho2131}.

\begin{figure}[ptbh]
\includegraphics[width=8.6cm]{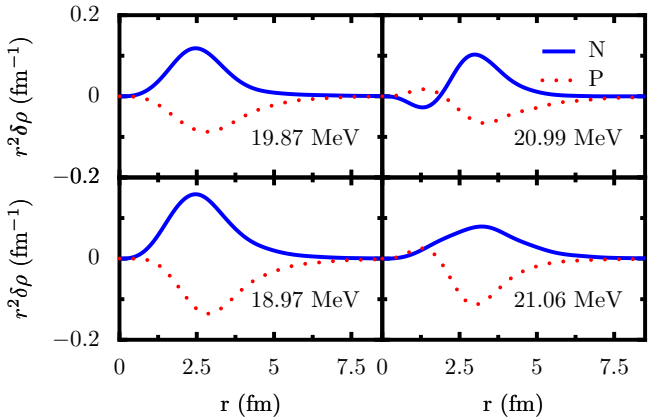}\caption{(Color online) Projected
transition densities for major $K^{\pi}=1^{-}$ peaks contributing to the
electric dipole response in $^{20}$Ne.}%
\label{fig:Ne20fourpeaks}%
\end{figure}

Coming back to Figure~\ref{fig:Ne20E1}, the response in the energy region
between 15~MeV and 25~MeV corresponds to the IVGDR. Its strength is heavily
fragmented into several peaks in an energy interval of about 3-4~MeV for both
excitation modes. The main contributions to the strength curve come from more
than four different peaks. For example, the $K^{\pi}=1^{-}$ IVGDR response is
composed, besides the already mentioned peak at 21.3~MeV, by four additional
major peaks, situated at 19.6, 20.2, 21.8, and 22.4~MeV respectively. Their
projected transition densities, in Figure~\ref{fig:Ne20fourpeaks}, show that
all of them can be classified as a vibration of the neutron density against
the proton density.

The decomposition in $ph$ components of the main $K^{\pi}=0^{-}$ and $K^{\pi
}=1^{-}$ IVGDR peaks can be found in Table~\ref{table1}. The characteristic
$\Delta N=1$ pattern of the IVGDR is present in both peaks. The high
collectivity indicates a very coherent excitation pattern that fits into the
properties of a giant resonance. This phenomenon can also be observed in the
transition densities, where the coherent superposition of $ph$ pairs is
evident in the absence of wavefunction-like features, and is easily
interpreted in a macroscopic picture where the proton and neutron densities
oscillate one against the other. The total percentage of the classical TRK
sum-rule exhausted between 10~MeV and 40~MeV for the calculated E1 response in
$^{20}$Ne is 111\%. In a fully classical system the share of the strength
exhausted by the $K^{\pi}=1^{-}$ mode should be double than that exhausted by
the $K^{\pi}=0^{-}$ mode; however, with 86\% of the TRK sum-rule coming from
the $K^{\pi}=1^{-}$ mode and 25\% from the $K^{\pi}=0^{-}$ mode, it is obvious
that it is no longer true for quantum systems, even though we do not fully
understand the mechanism behind this phenomenon.

\begin{table}[ptbh]
\begin{center}%
\begin{tabular}
[c]{lrlcrc}%
\multicolumn{5}{l}{$K^{\pi}=0^{-}$ peak at 16.73~MeV} & \\
&  &  &  &  & $\varepsilon_{1}-\varepsilon_{2}$\\\hline
N & 20\% & $\frac{3}{2}^{-}[101]$ & $-$ & $\frac{3}{2}^{+}[211]$ & 14.25\\
P & 18\% & $\frac{3}{2}^{-}[101]$ & $-$ & $\frac{3}{2}^{+}[211]$ & 13.89\\
N & 16\% & $\frac{1}{2}^{+}[220]$ & $-$ & $\frac{1}{2}^{-}[321]$ & 17.52\\
N & 11\% & $\frac{1}{2}^{-}[101]$ & $-$ & $\frac{1}{2}^{+}[200]$ & 15.28\\
P & 9\% & $\frac{1}{2}^{-}[101]$ & $-$ & $\frac{1}{2}^{+}[211]$ & 14.46\\
N & 7\% & $\frac{1}{2}^{+}[220]$ & $-$ & $\frac{1}{2}^{-}[310]$ & 18.20\\
P & 5\% & $\frac{1}{2}^{-}[101]$ & $-$ & $\frac{1}{2}^{+}[211]$ & 12.71\\
N & 3\% & $\frac{1}{2}^{-}[101]$ & $-$ & $\frac{1}{2}^{+}[211]$ & 13.37\\
N & 2\% & $\frac{1}{2}^{+}[220]$ & $-$ & $\frac{1}{2}^{-}[330]$ &
12.90\\\hline
\multicolumn{5}{l}{$K^{\pi}=1^{-}$ peak at 21.31~MeV} & \\
&  &  &  &  & $\varepsilon_{1}-\varepsilon_{2}$\\\hline
N & 13\% & $\frac{1}{2}^{-}[110]$ & $-$ & $\frac{3}{2}^{+}[211]$ & 20.01\\
N & 13\% & $\frac{1}{2}^{+}[220]$ & $-$ & $\frac{1}{2}^{-}[321]$ & 20.64\\
P & 11\% & $\frac{1}{2}^{+}[220]$ & $-$ & $\frac{1}{2}^{-}[321]$ & 22.03\\
P & 9\% & $\frac{1}{2}^{-}[101]$ & $-$ & $\frac{1}{2}^{+}[200]$ & 21.55\\
N & 7\% & $\frac{1}{2}^{-}[110]$ & $-$ & $\frac{1}{2}^{+}[211]$ & 24.96\\
P & 5\% & $\frac{1}{2}^{-}[101]$ & $-$ & $\frac{3}{2}^{+}[202]$ & 22.83\\
N & 5\% & $\frac{1}{2}^{+}[220]$ & $-$ & $\frac{1}{2}^{-}[321]$ & 23.81\\
N & 5\% & $\frac{1}{2}^{+}[220]$ & $-$ & $\frac{3}{2}^{-}[312]$ & 21.56\\
N & 4\% & $\frac{1}{2}^{+}[220]$ & $-$ & $\frac{3}{2}^{-}[321]$ &
22.79\\\hline
\end{tabular}
\end{center}
\caption{Particle-hole structure of the two main IVGDR modes. N and P indicate
a neutron or proton $ph$ pair, respectively. The second column is the
percentage of the contribution of each particular $ph$ excitation. The Nilsson
quantum numbers, labeling the particle-hole, are shown in the next columns.
The last column gives the energy of the excitation in MeV.}%
\label{table1}%
\end{table}

\section{Concluding remarks}

\label{sec:conclusions}

In the present investigation we have formulated the relativistic random phase
approximation (RRPA) on the basis of a relativistic mean-field (RMF) model
having axial symmetry in a fully self-consistent way, i.e., the interactions
used in both the RMF equations and in the matrix equation of the RRPA are
derived from the same Lagrangian, i.e. the same energy functional. As it has
been shown, this self-consistency feature is of vital importance for the
fulfillment of current conservation and the proper decoupling of spurious
modes without further adjustments in the interaction.

So far, pairing correlations have not been included. The inclusion of such
correlations will allow the application of this method to a large number of
investigations in medium and heavy nuclei, in particular, in a first step, for
a systematic study of low-lying electric and magnetic dipole strength over
large regions of the periodic table. Of course, the study of the nuclear
response to other electric and magnetic multipoles is also open to scrutiny.
Since the formulation of relativistic proton-neutron RPA, once the main
building blocks presented in this document are present, is mostly trivial, its
implementation opens the door to the wide area of nuclear spin-isospin
excitations, in particular the Isobaric Analog Resonance (IAR) and the
Gamov-Teller Resonance (GTR), but also for many types of weak processes such
as the $\beta$-decay and neutrino-reactions in deformed nuclei.

In conclusion, the relativistic RPA formulated for axially deformed systems
represents a significant new theoretical tool for a realistic description of
excitation phenomena in large regions of the nuclear chart, which has been
accessible so far only by relatively crude phenomenological models. Its
development, and the sample application presented in this document, show that
its future use in nuclear structure and astrophysics will provide an valuable
insight into very important, and still open, questions about the nature of
nuclear interaction, collective response, deformation effects and cross
sections relevant for astrophysical processes.

\begin{acknowledgments}
Helpful discussions with R. R. Hilton and D. Vretenar are gratefully
acknowledged. P.R. thanks for the support provided by the Ministerio de
Educaci\'on y Ciencia, Spain. The paper has also been supported by the
Bundesministerium f\"{u}r Bildung und Forschung, Germany under project 06 MT
246 and by the DFG cluster of excellence \textquotedblleft Origin and
Structure of the Universe\textquotedblright\ (www.universe-cluster.de).
\end{acknowledgments}

\appendix

\section{Two-body matrix elements}

\label{app:A} Starting form the general expression for the two-body matrix
element in Eq. (\ref{res:q-integ})
\begin{equation}
\langle kl^{\prime}|\hat{V}_{m}^{ph}|k^{\prime}l\rangle=\int\frac{d^{3}%
q}{(2\pi)^{3}}\langle k|\hat{Q}^{\mu}(\boldsymbol{q})|k^{\prime}\rangle
\Delta_{m}(\boldsymbol{q})\langle l|\hat{Q}_{\mu}(\boldsymbol{q})|l^{\prime
}\rangle^{\ast} \label{q-int}%
\end{equation}
we first have to evalute the matrix elements (\ref{spmatelm}) for the single
particle opreators $Q^{\mu}(\boldsymbol{q}).$ In cylindrical coordinates
\begin{equation}
\boldsymbol{q}=(q_{x},q_{y},q_{z})=(q\cos\chi,q\sin\chi,q_{z})
\end{equation}
we obtain
\begin{equation}
Q_{kk^{\prime}}^{\mu}(q,\chi,q_{z})=\int\frac{d\varphi}{2\pi}\,d^{2}%
r\,~\bar{\psi}_{k}g_{m}\,\Gamma^{\mu}\,\psi_{k^{\prime}}\,e^{iq_{z}%
z+iqr\cos(\varphi-\chi)} \label{r-int}%
\end{equation}
It turns out to be useful to \ classify the various vertices $\Gamma^{\mu}$ by
the spin quantum numbers $S$ and $S_{z}=\Sigma$ of the exchanged meson. For
this reason we use the the $\gamma$-matrices in the Dirac basis defined by
\begin{equation}
\{\gamma^{0},\gamma^{+}=\frac{-1}{\sqrt{2}}(\gamma^{1}+i\gamma^{2}),\gamma
^{-}=\frac{1}{\sqrt{2}}(\gamma^{1}-i\gamma^{2}),\gamma^{3}\}
\label{dirac-basis}%
\end{equation}
and obtain%
\begin{equation}
\gamma^{\mu}\gamma_{\mu}=\gamma^{0}\gamma^{0}+\gamma^{+}\gamma^{-}+\gamma
^{-}\gamma^{+}-\gamma^{3}\gamma^{3}%
\end{equation}
Including the scalar mesons (and neglecting for the moment the isospin) we
therefore have 5 vertices characterized by the index $\tilde{\mu}$:
\begin{equation}
\Gamma^{\tilde{\mu}}=(1,\gamma^{0},\gamma^{+},\gamma^{-},\gamma^{3}).
\end{equation}
$\tilde{\mu}$ runs over $\tilde{\mu}=s$ (for scaler mesons, $S=\Sigma=0$),
$\tilde{\mu}=0$ (for the time-like part of the vector mesons, $S=\Sigma=0$)
and $\tilde{\mu}=+,-,3$ (for the spatial parts of the vector mesons with
$S=1$). The channels $\tilde{\mu}=\pm$ describe the spin flip ($\Sigma=\pm1$)
and $\tilde{\mu}=3$ has $\Sigma=0$.

Using the Dirac spinors in cylindrical coordinates (\ref{def:wavef}) and
exploiting the properties of the $\gamma$-matrices defined in Eq.
(\ref{dirac-basis}), we find that the $\varphi$-dependence of the amplitudes
$\bar{\psi}_{k}(r,\varphi,z)\,\Gamma^{\tilde{\mu}}\,\psi_{k^{\prime}%
}(r,\varphi,z)$ can be separated
\begin{equation}
\bar{\psi}_{k}(r,\varphi,z)\,\Gamma^{\tilde{\mu}}\,\psi_{k^{\prime}}%
(r,\varphi,z)=i^{S}\,F_{kk^{\prime}}^{\tilde{\mu}}(r,z)e^{i\Lambda\varphi}%
\end{equation}
where the integer%
\begin{equation}
\Lambda=\Omega_{k^{{}}}-\Omega_{k^{\prime}}-\Sigma=K-\Sigma
\end{equation}
is the orbital part of the angular momentum of the pair in $z$-direction. The
real functions $F_{kk^{\prime}}^{\tilde{\mu}}(r,z)$ are given by
\begin{align}
F_{kk^{\prime}}^{s}(r,z)  &  =f_{k^{{}}}^{+}f_{k^{\prime}}^{+}+f_{k^{{}}}%
^{-}f_{k^{\prime}}^{-}-g_{k^{{}}}^{+}g_{k^{\prime}}^{+}-g_{k^{{}}}%
^{-}g_{k^{\prime}}^{-},\\
F_{kk^{\prime}}^{0}(r,z)  &  =f_{k^{{}}}^{+}f_{k^{\prime}}^{+}+f_{k^{{}}}%
^{-}f_{k^{\prime}}^{-}+g_{k^{{}}}^{+}g_{k^{\prime}}^{+}+g_{k^{{}}}%
^{-}g_{k^{\prime}}^{-},\\
F_{kk^{\prime}}^{+}(r,z)  &  =g_{k^{{}}}^{+}f_{k^{\prime}}^{-}-f_{k^{{}}}%
^{+}g_{k^{\prime}}^{-}\\
F_{kk^{\prime}}^{-}(r,z)  &  =f_{k^{{}}}^{-}g_{k^{\prime}}^{+}-g_{k^{{}}}%
^{-}f_{k^{\prime}}^{+}\\
F_{kk^{\prime}}^{3}(r,z)  &  =f_{k^{{}}}^{+}g_{k^{\prime}}^{+}-g_{k^{{}}}%
^{+}f_{k^{\prime}}^{+}+f_{k^{{}}}^{-}g_{k^{\prime}}^{-}-g_{k^{{}}}%
^{-}f_{k^{\prime}}^{-}%
\end{align}
This allows us to evaluate the $\varphi$-integration in the integral
(\ref{r-int}) analytically and to express it in terms of Bessel functions of
the first kind
\begin{equation}
J_{n}(x)=(-i)^{n}\int_{0}^{2\pi}\frac{d\varphi}{2\pi}\,e^{in\varphi}%
e^{ix\cos\varphi}.
\end{equation}
We obtain%
\begin{equation}
\mathrm{Q}_{kk^{\prime}}^{\tilde{\mu}}(\boldsymbol{q})=i^{\Lambda
+S}\,e^{i\Lambda\chi}\,\mathcal{F}_{kk^{\prime}}^{\tilde{\mu}}(q,q_{z})
\end{equation}
with the integrals
\begin{equation}
\mathcal{F}_{kk^{\prime}}^{\tilde{\mu}}(q,q_{z})=%
{\displaystyle\int}
\,d^{2}r\,\,F_{kk^{\prime}}^{\tilde{\mu}}(r,z)\,J_{\Lambda}(qr)\,e^{iq_{z}z}.
\end{equation}
which are either real or purely imaginary. Using the parity relation%
\begin{equation}
F_{kk^{\prime}}^{\tilde{\mu}}(r,-z)=\pi(-)^{S+\Lambda}F_{kk^{\prime}}%
^{\tilde{\mu}}(r,z)
\end{equation}
whe recognize that the exponential factor $e^{iq_{z}z}$ reduces to the cosine
or sine depending of the quantum numbers $\pi$ and $K$ of the mode and on the
spin quantum numbers $S,\Sigma$ of the vertex $\tilde{\mu}$%

\begin{equation}
e^{iq_{z}z}\rightarrow\left\{
\begin{array}
[c]{c}%
~\cos(q_{z}z)\text{ \ for \ \ }\pi(-)^{K+S-\Sigma}=+1\\
i\sin(q_{z}z)\text{ \ \ }\mathrm{for}\text{ \ \ }\pi(-)^{K+S-\Sigma}=-1
\end{array}
\right.
\end{equation}
Substitution of these expressions in the integral of Eq.~(\ref{q-int}) the
$\chi$-integration can be carried out analytically and leads to the selection
rule%
\begin{equation}
\Omega_{k}-\Omega_{k^{\prime}}=\Omega_{l}-\Omega_{l^{\prime}}.
\end{equation}
and to the following matrix elements for the exchange of scalar mesons%
\begin{equation}
\left\langle kl^{\prime}|\hat{V}_{\sigma}^{ph}|k^{\prime}l\right\rangle
=-\int\frac{d^{2}q}{(2\pi)^{2}}\mathcal{F}_{kk^{\prime}}^{s}\Delta_{\sigma
}\mathcal{F}_{ll^{\prime}}^{s\ast}, \label{E1}%
\end{equation}
of time-like part vector mesons%
\begin{equation}
\left\langle kl^{\prime}|\hat{V}_{\omega^{0}}^{ph}|k^{\prime}l\right\rangle
=+\int\frac{d^{2}q}{(2\pi)^{2}}\mathcal{F}_{kk^{\prime}}^{0}\Delta_{\omega
}\mathcal{F}_{ll^{\prime}}^{0\ast},
\end{equation}
and of space-like of the vector mesons%
\begin{align}
\left\langle kl^{\prime}|\hat{V}_{\mathbf{\omega}}^{ph}|k^{\prime
}l\right\rangle  &  =\int\frac{d^{2}q}{(2\pi)^{2}}\mathcal{F}_{kk^{\prime}%
}^{+}\Delta_{\omega}\mathcal{F}_{ll^{\prime}}^{-\ast}\label{E2}\\
&  +\int\frac{d^{2}q}{(2\pi)^{2}}\mathcal{F}_{kk^{\prime}}^{-}\Delta_{\omega
}\mathcal{F}_{ll^{\prime}}^{+\ast}\\
&  -\int\frac{d^{2}q}{(2\pi)^{2}}\mathcal{F}_{kk^{\prime}}^{3}\Delta_{\omega
}\mathcal{F}_{ll^{\prime}}^{3\ast},
\end{align}
where, for simplicity, we have neglected for each matrix element a factor
$\delta_{\Omega_{k^{{}}}-\Omega_{k^{\prime}},\Omega_{l^{{}}}-\Omega
_{l^{\prime}}}$ and the arguments in the functions $\mathcal{F}_{kk^{\prime}%
}^{\tilde{\mu}}(q,q_{z})$ and in the propagators%
\begin{equation}
\Delta_{m}(q,q_{z})=\frac{1}{q^{2}+q_{z}^{2}+m_{m}^{2}}.
\end{equation}
in the two-dimensional momentum integrals.

\section{Non-linear $\sigma$ propagator in momentum space}

\label{app:B} The equation to solve is
\begin{equation}
\lbrack-\Delta+m_{\sigma}^{2}+W(\boldsymbol{r})]\delta\sigma(\boldsymbol{r}%
)=-g_{\sigma}\delta\rho_{\text{s}}(\boldsymbol{r}) \label{B1}%
\end{equation}
with
\begin{equation}
W(\boldsymbol{r})=2g_{2}\sigma+3g_{3}\sigma^{2}%
\end{equation}
Because of axial symmetry and using cylindrical coordinates $\boldsymbol{r=}%
(r\cos\varphi,r\sin\varphi,z)$ $W(\boldsymbol{r}):=W(r,z)$ does not depend on
the azimuth angle $\varphi$. We solve Eq. (\ref{B1}) in momentum space.
$W(\boldsymbol{r})$ is local in $r$-space, but it is an operator in momentum
space%
\begin{equation}
W(\boldsymbol{q},\boldsymbol{q}^{\prime})=%
{\displaystyle\int}
d^{3}rW(\boldsymbol{r})e^{-i\boldsymbol{r}(\boldsymbol{q}-\boldsymbol{q}%
^{\prime})}%
\end{equation}
The propagator in momentum space is the solution of
\begin{equation}
(\boldsymbol{q}^{2}+m^{2})\Delta(\boldsymbol{q},\boldsymbol{q}^{\prime})+%
{\displaystyle\int}
d^{3}rW(\boldsymbol{q},\boldsymbol{q}^{\prime\prime})\Delta(\boldsymbol{q}%
^{\prime\prime},\boldsymbol{q}^{\prime})=\delta(\boldsymbol{q}-\boldsymbol{q}%
^{\prime}) \label{eq:nlprop}%
\end{equation}
where the $\ast$ is the convolution operator and $W(\boldsymbol{q})$ is the
Fourier transform of $W(\boldsymbol{r})$. Expanding the $\delta$-function in
cylindrical coordinates in $q$-space using $\boldsymbol{q=}(q\cos\chi
,q\sin\chi,q_{z})$ we find
\begin{equation}
\delta(\boldsymbol{q}-\boldsymbol{q}^{\prime})=\frac{\delta(q-q^{\prime})}%
{q}\delta(q_{z}-q_{z}^{\prime})\sum_{n=-\infty}^{\infty}e^{in(\chi
-\chi^{\prime})}%
\end{equation}
and taking the following ansatz for $\Delta$
\begin{equation}
\Delta(\boldsymbol{q},\boldsymbol{q}^{\prime})=\sum_{n=-\infty}^{\infty}%
\Delta_{n}(q,q_{z},q^{\prime},q_{z}^{\prime})\;e^{in(\chi-\chi^{\prime})}%
\end{equation}
and inserting it in Eq.~\eqref{eq:nlprop} leads to a set of integral equations
for each $\Delta_{n}$
\begin{align}
&  (q_{{}}^{2}+q_{z}^{2}+m^{2})\Delta_{n}(q,q_{z},q^{\prime},q_{z}^{\prime
})\nonumber\\
&  +\int d^{2}q^{\prime\prime}W_{n}(q,q_{z},q^{\prime\prime},q_{z}%
^{\prime\prime})\Delta_{n}(q^{\prime\prime},q_{z}^{\prime\prime},q^{\prime
},q_{z}^{\prime})\nonumber\\
&  =\frac{\delta(q-q^{\prime})}{q}\delta(q_{z}-q_{z}^{\prime}),
\label{eq:deltasnl}%
\end{align}
where we have used the obvious notation $d^{2}q=qdqdq_{z}$. Each $W_{n}$ can
be calculated using the following series expansion
\begin{equation}
e^{\mathrm{i}xcos(\alpha)}=\sum_{n=-\infty}^{\infty}\mathrm{i}^{n}%
J_{n}(x)e^{\mathrm{i}n\alpha}%
\end{equation}
that leads to the following expression for the non-linear $\sigma$ field in
momentum space
\begin{equation}
W_{n}(q,q_{z},q^{\prime},q_{z}^{\prime})=\int\frac{d^{2}r}{2\pi}%
W(r,z)e^{-i(q_{z}-q_{z}^{\prime})z}J_{n}(qr)J_{n}(q^{\prime}r)\nonumber
\end{equation}
that together with Eq.~\ref{eq:deltasnl} allow the numerical evaluation of the
non-linear $\sigma$ propagator in momentum space.

\section{Evaluation of the M1 single particle matrix elements}

\label{app:C} The M1 operator is defined as:
\begin{equation}
\hat{M}_{1\mu}=\sqrt{\frac{3}{4\pi}}\mu_{N}\left(  g_{s}\boldsymbol{s}%
+g_{l}\boldsymbol{l}\right)
\end{equation}
with
\begin{equation}%
\begin{array}
[c]{ccc}%
g_{s} & = & g_{p}\\
g_{l} & = & 1
\end{array}
\}\quad\mathrm{protons}\qquad%
\begin{array}
[c]{ccc}%
g_{s} & = & g_{n}\\
g_{l} & = & 0
\end{array}
\}\quad\mathrm{neutrons}%
\end{equation}
In the spherical coordinates defined as:
\begin{equation}
x_{+}=\frac{-1}{\sqrt{2}}\left(  x+iy\right)  ,\text{ \ \ }x_{-}=\frac
{1}{\sqrt{2}}\left(  x-iy\right)  ,\text{ \ \ }x_{0}=z \label{sphep1def}%
\end{equation}
we find%
\begin{align}
s_{+}  &  =\frac{1}{2}\Sigma_{+}=\frac{-1}{2\sqrt{2}}\left(  \Sigma
_{x}+i\Sigma_{y}\right) \\
l_{+}  &  =\frac{\mathds{1}}{2\sqrt{2}}e^{i\varphi}\left[  r\partial
_{z}-z\left(  \partial_{r}+i\frac{1}{r}\partial_{\varphi}\right)  \right]
\end{align}
and in the $\hat{M}_{11}$ single particle matrix elements the integration over
the azimuthal angle $\varphi$ can be carried out analytically. This yields
\begin{align}
\langle k|\hat{M}_{11}|k^{\prime}\rangle &  =\mu_{N}\delta_{\Omega_{m}%
-\Omega_{i},1}\frac{1}{\sqrt{2}}\sqrt{\frac{3}{4\pi}}\int d^{2}%
rg_{l} \nonumber \\
\times \biggl[
&  +r\left(  f_{k^{{}}}^{+}\partial_{z}f_{k^{\prime}}^{+}+f_{k^{{}}}%
^{-}\partial_{z}f_{k^{\prime}}^{-}+g_{k^{{}}}^{+}\partial_{z}g_{k^{\prime}%
}^{+}+g_{k^{{}}}^{-}\partial_{z}g_{k^{\prime}}^{-}\right) \nonumber\\
&  -z\left(  f_{k^{{}}}^{+}\partial_{r}f_{k^{\prime}}^{+}+f_{k^{{}}}%
^{-}\partial_{r}f_{k^{\prime}}^{-}+g_{k^{{}}}^{+}\partial_{r}g_{k^{\prime}%
}^{+}+g_{k^{{}}}^{+}\partial_{r}g_{k^{\prime}}^{-}\right) \nonumber\\
&  +\frac{(\Omega_{i}-\frac{1}{2})z}{r}(f_{k^{{}}}^{+}f_{k^{\prime}}%
^{+}+g_{k^{{}}}^{+}g_{k^{\prime}}^{+})\nonumber\\
&  +\frac{(\Omega_{i}+\frac{1}{2})z}{r}(f_{k^{{}}}^{-}f_{k^{\prime}}%
^{-}+g_{k^{{}}}^{-}g_{k^{\prime}}^{-})\nonumber\\
&  -g_{s}\left(  f_{k^{{}}}^{+}f_{k^{\prime}}^{-}+g_{k^{{}}}^{+}g_{k^{\prime}%
}^{-}\right)\biggr]
\end{align}

\section{Evaluation of the E1 single particle matrix elements}

\label{app:D} The effective isovector dipole operator, with spurious
translation of the center of mass already subtracted, reads in spherical
coordinates
\begin{equation}
\boldsymbol{\hat{D}}=e\frac{N}{A}\sum_{p=1}^{Z}\boldsymbol{r}_{p}-e\frac{Z}%
{A}\sum_{n=1}^{N}\boldsymbol{r}_{n}%
\end{equation}
With the spherical coordinates of Eq.~\eqref{sphep1def} the dipole operators
are given in cylindrical coordinates as
\begin{align}
\hat{D}_{0}  &  =e\frac{N}{A}\sum_{p=1}^{Z}z_{p}-e\frac{Z}{A}\sum_{n=1}%
^{N}z_{n},\\
\hat{D}_{\pm}  &  =e\frac{N}{A}\sum_{p=1}^{Z}r_{p}e^{\pm i\varphi_{p}}%
-e\frac{Z}{A}\sum_{n=1}^{N}r_{n}e^{\pm i\varphi_{n}},
\end{align}
and the single particle matrix elements are
\begin{align}
\langle k|\hat{D}_{0}|k_{0}^{\prime}\rangle &  =e_{\text{eff}}~\delta
_{\Omega_{k^{{}}},\Omega_{k^{\prime}}}\int zd^{2}r\\
&  \text{ \ \ \ \ }\left(  f_{k^{{}}}^{+}f_{k^{\prime}}^{+}+f_{k^{{}}}%
^{-}f_{k^{\prime}}^{-}+g_{k^{{}}}^{+}g_{k^{\prime}}^{+}+g_{k^{{}}}%
^{-}g_{k^{\prime}}^{-}\right) \nonumber\\
\langle k|\hat{D}_{+}|k_{0}^{\prime}\rangle &  =e_{\text{eff}}~\delta
_{\Omega_{k^{{}}},\Omega_{k^{\prime}}+1}\int rd^{2}r\\
&  \text{ \ \ \ \ }\left(  f_{k^{{}}}^{+}f_{k^{\prime}}^{+}+f_{k^{{}}}%
^{+}f_{k^{\prime}}^{+}+g_{k^{{}}}^{+}g_{k^{\prime}}^{+}+g_{k^{{}}}%
^{+}g_{k^{\prime}}^{+}\right) \nonumber
\end{align}
where $e_{\text{eff}}=eN/A$ for proton pairs and $e_{\text{eff}}=-eZ/A$ for
neutron pairs.

\bigskip

\section{Approximate angular momentum projection}

\label{app:E}

The wave function $\vert\psi_{IM}\rangle$ in the laboratory frame is
obtained by angular momentum projection from the intrinsic wave
function $\vert\Phi\rangle$
\begin{equation}
\vert\psi_{IM}\rangle=\sum_{K}g_{K}^{I}\hat{P}_{MK}^{I}\vert\Phi\rangle
\end{equation}
where the projector operator $\hat{P}_{MK}^{I}$ is given by \cite{Ed.57}
\begin{equation}
\hat{P}_{MK}^{I}=\frac{2I+1}{8\pi^{2}}\int d\Omega\mathcal{D}_{MK}^{I\star
}(\Omega)\hat{R}(\Omega).
\end{equation}
$\Omega$ represents the Euler angles ($\alpha$,$\beta$,$\gamma$),
$\mathcal{D}_{MK}^{I}(\Omega)$ are the Wigner functions \cite{VMK.88}
and
$\hat{R}(\Omega)=e^{-i\alpha\hat{J}_{z}}e^{-i\beta\hat{J}_{y}}e^{-i\gamma
\hat{J}_{z}}$ is the rotation operator. Taking into account the
transformation law for the multipole operators $\hat{Q}_{\lambda\mu}$
under rotations
\begin{equation}
\hat{R}(\Omega)\hat{Q}_{\lambda\mu}\hat{R}^{\dagger}(\Omega)=\sum_{\mu
^{\prime}}\mathcal{D}_{\mu^{\prime}\mu}^{\lambda}(\Omega)\hat{Q}_{\lambda
\mu^{\prime}}%
\end{equation}
The matrix element of this operator between two states with good angular
momentum is given by
\begin{equation}
\langle \Psi_{I_{f}M_{f}} \vert \hat{Q}_{\lambda\mu} \vert
\Psi_{I_iM_i}\rangle =%
\frac {\langle I_{i}M_{i}\lambda\mu\vert I_{f}M_{f}\rangle}
{\sqrt{2I_{f}+1}}%
\langle I_f \vert\vert \hat{Q}_{\lambda} \vert\vert I_i\rangle
\end{equation}
with the reduced matrix element defined by
\begin{align}
\langle I_fK_f&\vert\vert\hat{Q}_\lambda\vert\vert I_iK_i\rangle%
=\frac{(2I_{i}+1)(2I_{f}+1)}{8\pi^{2}}(-)^{I_{i}-\lambda}\\%
&\times  \sum_{\substack{K_{i},K_{f}\\\mu,\mu^{\prime}}}(-)^{K_{f}}g_{K_{f}%
}^{I_{f}\star}g_{K_{i}}^{I_{i}}%
\left(
\begin{array}
[c]{ccc}%
I_{i} & \lambda & I_{f}\\
\mu^{\prime} & \mu & -K_{f}%
\end{array}
\right)%
\nonumber\\%
&~~~~~~~\times%
\int d\Omega~\mathcal{D}_{\mu^{\prime}K_{i}}^{I_{i}\star}%
(\Omega )\langle K_{f} \vert \hat{Q}_{\lambda\mu} \hat{R}(\Omega)
\vert K_{i} \rangle \nonumber
\end{align}
In the case of axial symmetry, the integral in the last equation is
reduced to
\begin{equation}
\int_{0}^{\pi}d(\cos\beta)~d_{-\mu^{\prime}K_{i}}^{I_{i}\star}(\beta
)\langle K_{f}\vert \hat{Q}_{\lambda\mu}e^{-i\beta \hat{J}_{y}} \vert
K_{i} \rangle
\end{equation}
To evaluate the overlap integrals in the last equation we use in the limit of
the needle approximation \cite{Vil.66,Zeh.67} $\ $to first order in a Kamlah
\cite{RS.80} expansion
\begin{equation}
\langle K_{f}\vert \hat{Q}_{\lambda\mu}e^{-i\beta \hat{J}_{y}} \vert
K_{i} \rangle=%
\langle K_{f}\vert \hat{Q}_{\lambda\mu} \vert K_{i}\rangle%
\langle K_{f}\vert e^{-i\beta \hat{J}_{y}} \vert K_{f} \rangle
\end{equation}
and using that the integral over $\beta$ contributes only at $\beta=0$ and
$\pi$ we obtain the final expression for approximate angular momentum
projection used in the calculation of RPA single-particle observables:
\begin{eqnarray}
&\langle I_f K_f||\hat{\mathcal{O}}_\lambda||I_iK_i\rangle=%
(2I_i+1)(2I_f+1)~~~~~~~~~~~~~~~~~~~
\\
&\times \left[%
\left(
\begin{array}[c]{ccc}%
I_{i} & \lambda & I_{f}\\
K_{i} & \mu & K_{f}%
\end{array}
\right)%
\langle K_f|\hat{\mathcal{O}}_{\lambda\mu}|K_i\rangle%
\right.~~~~~~~~~~~~
\nonumber\\
&  \left.  +(-1)^{I_i+K_i}%
\left(
\begin{array}[c]{ccc}%
I_i & \lambda & I_f\\
\bar{K}_i & \mu & K_f%
\end{array}
\right)%
\langle K_f|\hat{\mathcal{O}}_{\lambda\mu}|\bar{K}_i\rangle\right]%
\nonumber%
\label{app:redmatelm}%
\end{eqnarray}


\begin{thebibliography}{99}                                                                                               %


\bibitem {HK.64}P. Hohenberg and W. Kohn, Phys. Rev. \textbf{136}, B864 (1964).

\bibitem {KS.65}W. Kohn and L.~J. Sham, Phys. Rev. \textbf{137}, A1697 (1965).

\bibitem {Eng.07}J. Engel, Phys. Rev. \textbf{C75}, 014306 (2007).

\bibitem {Gir.08}R.~G. Giraud, (2008), arXiv:0801.3447[nucl-th].

\bibitem {BHR.03}M. Bender, P.-H. Heenen, and P.-G. Reinhard, Rev. Mod. Phys.
\textbf{75}, 121 (2003).

\bibitem {Rin.96}P. Ring, Prog. Part. Nucl. Phys. \textbf{37}, 193 (1996).

\bibitem {VALR.05}D. Vretenar, A.~V. Afanasjev, G.~A. Lalazissis, and P. Ring,
Phys. Rep. \textbf{ 409}, 101 (2005).

\bibitem {GEL.96}T. Gonzales-Llarena, J.~L. Egido, G.~A. Lalazissis, and P.
Ring, Phys. Lett. \textbf{B379}, 13 (1996).

\bibitem {BB.77}J. Boguta and A.~R. Bodmer, Nucl. Phys. \textbf{A292}, 413 (1977).

\bibitem {VBR.95}D. Vretenar, H. Berghammer, and P. Ring, Nucl. Phys.
\textbf{A581}, 679 (1995).

\bibitem {RS.80}P. Ring and P. Schuck, \emph{The nuclear many-body problem}
(Springer, Heidelberg, 1980).

\bibitem {RMG.01}P. Ring, Z.-Y. Ma, N. Van~Giai, D. Vretenar, A. Wandelt, and
L.-G. Cao, Nucl. Phys. \textbf{A694}, 249 (2001).

\bibitem {Fur.85}R.~J. Furnstahl, Phys. Lett. \textbf{B152}, 313 (1985).

\bibitem {MWG.02}Z.-Y. Ma, A. Wandelt, N. Van~Giai, D. Vretenar, P. Ring, and
L.-G. Cao, Nucl. Phys. \textbf{A703}, 222 (2002).

\bibitem {PRB.87}W. Pannert, P. Ring, and J. Boguta, Phys. Rev. Lett.
\textbf{59}, 2420 (1987).

\bibitem {GRT.90}Y.~K. Gambhir, P. Ring, and A. Thimet, Ann. Phys. (N.Y.)
\textbf{198}, 132 (1990).

\bibitem {HHR.98}R. Hilton, W. H\"{o}henberger, and P. Ring, Euro. Phys. J
\textbf{A1}, 257 (1998).

\bibitem {NKK.06}V. Nesterenko, W. Kleinig, J. Kvasil, P. Vesely, and P.-G.
Reinhard, (2006), arXiv:0610040.v1[nucl-th].

\bibitem {PGB.07}S. Peru, H. Goutte, and J.~F. Berger, Nucl. Phys.
\textbf{A788}, 44c (2007).

\bibitem {Gor.98}S. Goriely, Phys. Lett. \textbf{B436}, 10 (1998).

\bibitem {VB.72}D. Vautherin and D.~M. Brink, Phys. Rev. \textbf{C5}, 626 (1972).

\bibitem {DG.80}J. Decharg\'{e} and D. Gogny, Phys.Rev. \textbf{C21}, 1568 (1980).

\bibitem {Wal.74}J.~D. Walecka, Ann. Phys. (N.Y.) \textbf{83}, 491 (1974).

\bibitem {Bru.54}K.~A. Brueckner, Phys. Rev. \textbf{97}, 508 (1954).

\bibitem {BT.92}R. Brockmann and H. Toki, Phys. Rev. Lett. \textbf{68}, 3408 (1992).

\bibitem {NJL.61a}Y. Nambu and G. Jona-Lasinio, Phys. Rev. \textbf{122}, 345 (1961).

\bibitem {BMM.02}T. B\"{u}rvenich, D.~G. Madland, J.~A. Maruhn, and P.-G.
Reinhard, Phys. Rev. \textbf{C65}, 044308 (2002).

\bibitem {TW.99}S. Typel and H.~H. Wolter, Nucl. Phys. \textbf{A656}, 331 (1999).

\bibitem {DD-ME1}T. Nik{\v{s}}i{\'{c}}, D. Vretenar, P. Finelli, and P. Ring,
Phys. Rev. \textbf{ C66}, 024306 (2002).

\bibitem {DD-ME2}G.~A. Lalazissis, T. Nik{\v{s}}i{\'{c}}, D. Vretenar, and P.
Ring, Phys. Rev. \textbf{C71}, 024312 (2005).

\bibitem {Bod.91}A.~R. Bodmer, Nucl. Phys. \textbf{A526}, 703 (1991).

\bibitem {TM1}Y. Sugahara and H. Toki, Nucl. Phys. \textbf{A579}, 557 (1994).

\bibitem {SFM.00}M.~M. Sharma, A.~R. Farhan, and S. Mythili, Phys. Rev.
\textbf{C61}, 054306 (2000).

\bibitem {HP.01}C.~J. Horowitz and J. Piekarewicz, Phys. Rev. \textbf{C64},
062802 (2001).

\bibitem {NL3}G.~A. Lalazissis, J. K\"{o}nig, and P. Ring, Phys. Rev.
\textbf{C55}, 540 (1997).

\bibitem {KS.65a}W. Kohn and L.~J. Sham, Phys. Rev. \textbf{140}, A1133 (1965).

\bibitem {CW.74}S.~A. Chin and J.~D. Walecka, Phys. Lett. \textbf{B52}, 24 (1974).

\bibitem {Chin.77}S.~A. Chin, Ann. Phys. (N.Y.) \textbf{108}, 301 (1977).

\bibitem {HS.84}C.~J. Horowitz and B.~D. Serot, Phys. Lett. \textbf{B140}, 181 (1984).

\bibitem {BS.84}A. Bielajew and B. Serot, Ann. Phys. (N.Y.) \textbf{156}, 215 (1984).

\bibitem {Per.87}R.~J. Perry, Nucl. Phys. \textbf{467}, 717 (1987).

\bibitem {HEM.00}T. Hartmann, J. Enders, P. Mohr, K. Vogt, S. Volz, and A.
Zilges, Phys. Rev. Lett. \textbf{85}, 274 (2000).

\bibitem {Was.88}D.~A. Wasson, Phys. Lett. \textbf{B210}, 41 (1988).

\bibitem {ZMR.91}Z.~Y. Zhu, H.~J. Mang, and P. Ring, Phys. Lett \textbf{B254},
325 (1991).

\bibitem {DF.90}J.~F. Dawson and R.~J. Furnstahl, Phys. Rev. \textbf{C42},
2009 (1990).

\bibitem {MGT.97}Z.-Y. Ma, N. Van~Giai, H. Toki, and M. L'Huillier, Phys. Rev.
\textbf{C42}, 2385 (1997).

\bibitem {Pap.07}P. Papakonstantinou, Euro. Phys. Lett. \textbf{78}, 12001 (2007).

\bibitem {Boh.36}N. Bohr, Nature \textbf{137}, 344 (1936).

\bibitem {Rai.50}J. Rainwater, Phys. Rev. \textbf{79}, 432 (1950).

\bibitem {Boh.51}A. Bohr, Phys. Rev. \textbf{81}, 134 (1951).

\bibitem {Boh.52}A. Bohr, Mat. Fys. Medd. Dan. Vid. Selsk. \textbf{26}, No. 14 (1952).

\bibitem {Ed.57}A.~R. Edmonds, \emph{Angular Momentum in Quantum Mechanics}
(University Press, Princeton, 1957).

\bibitem {KST.04}S.~P. Kamerdzhiev, J. Speth, and G.~Y. Tertychny, Phys. Rep.
\textbf{393}, 1 (2004).

\bibitem {LR.06}E. Litvinova and P. Ring, Phys. Rev. \textbf{C73}, 044328 (2006).

\bibitem {LRT.07}E. Litvinova, P. Ring, and V.~I. Tselyaev, Phys. Rev.
\textbf{C75}, 064308 (2007).

\bibitem {PW.85}C.~E. Price and G.~E. Walker, Phys. Lett. \textbf{B155}, 17 (1985).

\bibitem {SRM.89}J.~R. Shepard, E. Rost, and J.~A. McNeil, Phys. Rev.
\textbf{C40}, 2320 (1989).

\bibitem {MFR.89}J.~A. McNeil, R.~J. Furnstahl, E. Rost, and J. Shepard, Phys.
Rev. \textbf{C40}, 399 (1989).

\bibitem {TV.62}D.~J. Thouless and J.~G. Valatin, Nucl. Phys. \textbf{31}, 211 (1962).

\bibitem {Tho.61}D.~J. Thouless, Nucl. Phys. \textbf{22}, 78 (1961).

\bibitem {Mar.77b}E.~R. Marshalek, Nucl. Phys. \textbf{A275}, 416 (1977).

\bibitem {BRS.84}D. Bohle, A. Richter, W. Stephen, A.~E.~L. Dieperink, N.
lo~Iudice, F. Palumbo, and O. Scholten, Phys. Lett. \textbf{B137}, 27 (1984).

\bibitem {SR.77}T. Suzuki and D.~J. Rowe, Nucl. Phys. \textbf{A286}, 307 (1977).

\bibitem {IP.78a}N. lo~Iudice and F. Palumbo, Phys. Rev. Lett. \textbf{41},
1532 (1978).

\bibitem {IP.78b}N. lo~Iudice and F. Palumbo, Phys. Rev. Lett. \textbf{74},
1046 (1978).

\bibitem {Hil.84}R.~R. Hilton, Z. Phys. \textbf{A316}, 121 (1984).

\bibitem {Hil.92}R.~R. Hilton, Ann. Phys. (N.Y.) \textbf{214}, 258 (1992).

\bibitem {Iac.81}F. Iachello, Nucl. Phys. \textbf{A358}, 89c (1981).

\bibitem {Die.83}A.~E.~L. Dieperink, in \emph{Prog. Part. Nucl. Phys.}, edited
by D.~H. Wilkinson (Pergamon Press, Oxford, 1983), Vol.~9, p.\ 121.

\bibitem {Iac.84}F. Iachello, Phys. Rev. Lett. \textbf{53}, 1427 (1984).

\bibitem {VHJ.86}P.~V. Isacker, K. Heyde, J. Jolie, and A. Sevrin, Ann. Phys.
(N.Y.) \textbf{171}, 253 (1986).

\bibitem {KPZ.96}U. Kneissl, H. Pitz, and A. Zilges, Progr. Part. Nucl. Phys.
\textbf{37}, 349 (1996).

\bibitem {Rich.95}A. Richter, Progr. Part. Nucl. Phys. \textbf{34}, 261 (1995).

\bibitem {Zaw.98}D. Zawischa, Nucl. Part. Phys. \textbf{24}, 683 (1998).

\bibitem {BalS.07a}E.~B. Balbutsev and P. Schuck, Ann. Phys. (N.Y.)
\textbf{322}, 489 (2007).

\bibitem {BalS.07b}E.~B. Balbutsev and P. Schuck, (2007), arXiv:\-0701039\-[nucl-th].

\bibitem {FWR.90}D. Frekers, H. Wortche, and A. Richter, \textit{et al.},
Phys. Lett. \textbf{B244}, 170 (1990).

\bibitem {CP.86}L. Chaves and A. Poves, Phys. Rev. \textbf{C34}, 1137 (1986).

\bibitem {LZ.87}H. Liu and L. Zamick, Phys. Rev. \textbf{C36}, 2057 (1987).

\bibitem {Dan.58}M. Danos, Nucl. Phys. \textbf{5}, 23 (1958).

\bibitem {Oka.58}K. Okamoto, Phys. Rev. \textbf{110}, 143 (1958).

\bibitem {VMK.88}D.~A. Varshalovich, A.~N. Moskalev, and V.~K. Khersonskii,
\emph{Quantum theory of angular momentum} (World Scientific Publishing Co.,
Inc., New Jersey, 1988).

\bibitem {Vil.66}F. Villars, Varenna Lectures \textbf{36}, 1; 14 (1966).

\bibitem {Zeh.67}H.~D. Zeh, Z. Phys. \textbf{202}, 28 (1967).
\end{thebibliography}

\end{document}